\documentclass[prd,twocolumn,superscriptaddress,altaffilletter,nofootinbib]{revtex4}
\usepackage[dvips]{graphicx}
\usepackage{amsmath}
\usepackage{graphicx,epsfig}
\newcommand{\be}{\begin{equation}}
\newcommand{\ee}{\end{equation}}
\newcommand{\bea}{\begin{eqnarray}}
\newcommand{\eea}{\end{eqnarray}}
\newcommand{\der}{\partial}

\begin{document}

%%%%%%%%%%%%%%%%%%%%%%%%%%%%%%%%%%%%%%%%%%%%%%%%%%%%%%%%%%%%%%%%%%%%%%%%%%%%%%%%%%%%%%%%%%%%%%%%%%%%%%

\title{On the quantum origin of inflation in the geometric inflation model}

%-------------------------------------------------------------------------------------------------------------------------------------------

\author{Israel Quiros}\email{iquiros@fisica.ugto.mx}\affiliation{Dpto. Ingenier\'ia Civil, Divisi\'on de Ingenier\'ia, Universidad de Guanajuato, Gto., CP 36000, M\'exico.}

\author{Roberto De Arcia}\email{robertodearcia@gmail.com}\affiliation{Depto. de Astronom\'ia, Divisi\'on de Ciencias Exactas, Universidad de Guanajuato, Gto., CP 36240, M\'exico.}

\author{Ricardo Garc\'ia-Salcedo}\email{rigarcias@ipn.mx}\affiliation{CICATA-Legaria, Instituto Polit\'ecnico Nacional, Ciudad de México, CP 11500, México.}

\author{Tame Gonzalez}\email{tamegc72@gmail.com}\affiliation{Dpto. Ingenier\'ia Civil, Divisi\'on de Ingenier\'ia, Universidad de Guanajuato, Gto., CP 36000, M\'exico.}

\author{Francisco X. Linares Cede\~no}\email{francisco.linares@umich.mx}\affiliation{Instituto de F\'isica y Matem\'aticas, Universidad Michoacana de San Nicol\'as de Hidalgo, Edificio C-3, Ciudad Universitaria, CP 58040 Morelia, Michoac\'an, M\'exico.}

\author{Ulises Nucamendi}\email{unucamendi@gmail.com}\affiliation{Instituto de F\'isica y Matem\'aticas, Universidad Michoacana de San Nicol\'as de Hidalgo, Edificio C-3, Ciudad Universitaria, CP 58040 Morelia, Michoac\'an, M\'exico.}\affiliation{Departamento de F\'isica, Cinvestav, Avenida Instituto Polit\'ecnico Nacional 2508, San Pedro Zacatenco, 07360, Gustavo A. Madero, Ciudad de M\'exico, M\'exico.}

\date{\today}

%-----------------------------------------------------------------------------

\begin{abstract} In this paper we investigate the cosmological dynamics of geometric inflation by means of the tools of the dynamical systems theory. We focus in the study of two explicit models where it is possible to sum the infinite series of higher curvature corrections that arises in the formalism. These would be very interesting possibilities since, if regard gravity as a quantum effective theory, a key feature is that higher powers of the curvature invariants are involved at higher loops. Hence, naively, consideration of the whole infinite tower of curvature invariants amounts to consideration of all of the higher order loops. The global dynamics of these toy models in the phase space is discussed and the quantum origin of primordial inflation is exposed.
\end{abstract}

%\pacs{02.30.Hq, 04.50.Kd, 05.45.-a, 47.10.Fg, 98.80.-k}

\maketitle

%%%%%%%%%%%%%%%%%%%%%%%%%%%%%%%%%%%%%%%

\section{Introduction}\label{sec-intro}

Since long ago it has been suggested that the quantum gravity action should contain, in addition to the Einstein-Hilbert action, contributions from higher-order curvature invariants involving more than the first two derivatives of the metric tensor \cite{utiyama}. Although such higher-derivative terms in the action would carry negligible consequences in the classical (infra-red) domain, at high frequencies they would dominate, leading to power-counting renormalizability \cite{veltman, dewitt}. It has been shown that gravitational actions which include terms quadratic in the curvature tensor are indeed renormalizable although not unitary \cite{stelle}. After due consideration of the Gauss-Bonnet topological invariant, 

\bea {\cal X}_4=R^2-4R_{\mu\nu}R^{\mu\nu}+R_{\mu\nu\lambda\sigma}R^{\mu\nu\lambda\sigma},\label{gb-inv}\eea only terms $\propto R^2$ and $\propto R_{\mu\nu}R^{\mu\nu}$, were considered in \cite{stelle}. As shown in \cite{stelle-1} (see also \cite{hindawi, hindawi-1}) such a quadratic (two-parametric) class of theories yields to a class of multimass models of gravity with a total of eight degrees of freedom: In addition to the usual massless excitations of the field (2 degrees of freedom), there are now massive spin-two (5 degrees of freedom) and a massive scalar excitations. The massive spin-two part of the field has negative energy so that it is a ghost excitation which, although for an effective theory could not be catastrophic, from a quantum perspective leads to non-unitarity. 

In four-dimensional space the class of theories derivable from Lagrangians that depend exclusively on the metric tensor field ${\cal L}={\cal L}(g_{\mu\nu})$ (and of its derivatives), and that admit 2nd order equations of motion, is limited by the Lovelock theorem \cite{lovelock, lovelock-1}. According to this theorem the only 2nd order Euler-Lagrange equation ${\cal E}_{\mu\nu}=0$, obtainable in a four-dimensional space from a Lagrangian of the form ${\cal L}(g_{\mu\nu})$ is when \cite{clifton-rev}: ${\cal E}_{\mu\nu}=\sqrt{|g|}\left(\alpha G_{\mu\nu}+\lambda g_{\mu\nu}\right),$ where $\alpha$, and $\lambda$ are constants and $G_{\mu\nu}=R_{\mu\nu}-g_{\mu\nu}R/2$ is the Einstein's tensor. The general class of Lagrangians that lead to these 2nd order equations of motion reads:

\bea {\cal L}=\sqrt{|g|}\left(\alpha R+\beta{\cal X}_4\right)+\gamma\epsilon^{\mu\nu\sigma\lambda}R^{\alpha\beta}_{\;\;\;\mu\nu}R_{\alpha\beta\sigma\lambda},\nonumber\eea where the second and third terms above do not contribute to the Euler-Lagrange equations. As a consequence, in four dimensions, the only viable alternatives to general relativity (GR) should be based either i) on the consideration of other fields beyond the metric or ii) on the assumption of higher-dimensional spaces, or iii) on the inclusion of higher than second derivatives of the metric in the equations of motion, among a few others \cite{clifton-rev}.

Hence, an interesting possibility to evade the Lovelock theorem in order to obtain equations of motion that differ from the GR ones, is to relax the requirement of considering up to 2nd-order derivatives of the metric. This has been, precisely, the route followed in a series of recent works \cite{bueno-prd-2016, bueno-prd-2016-1, mann-prd-2017, chinese, arciniega-plb-2020, oliva, arciniega-2}, in order to obtain dynamics substantially different from those resulting from GR.

%----------------------ECG------------------------------

Among these ``beyond Lovelock'' proposals is the so called Einsteinian cubic gravity (ECG) theory \cite{bueno-prd-2016, bueno-prd-2016-1, mann-prd-2017}. The ECG formalism is the outcome of an approach based on a $D$-dimensional theory involving arbitrary contractions of the Riemann tensor and the metric:

\bea S=\int d^Dx\sqrt{|g|}{\cal L}(g_{\mu\nu},R_{\mu\nu\sigma\lambda}),\label{pablo-lag}\eea whose motion equations

\bea {\cal E}_{\mu\nu}=P_{\mu\sigma\rho\lambda}R_\nu^{\;\sigma\rho\lambda}-\frac{1}{2}\,g_{\mu\nu}{\cal L}-2\nabla^\lambda\nabla^\sigma P_{\mu\lambda\sigma\nu}=0,\label{pablo-feq}\eea where ${\cal E}_{\mu\nu}$ is the Euler–Lagrange tensor, and

\bea\left. P^{\mu\nu\sigma\lambda}\equiv\frac{\der{\cal L}}{\der R_{\mu\nu\sigma\lambda}}\right|_{g_{\alpha\beta}},\nonumber\eea contain up to fourth-order derivatives of the metric. The linearization of \eqref{pablo-feq} around maximally symmetric backgrounds with Riemann tensor $$R^{(0)}_{\mu\nu\sigma\lambda}=\Lambda\left[g^{(0)}_{\mu\sigma}g^{(0)}_{\lambda\nu}-g^{(0)}_{\mu\lambda}g^{(0)}_{\sigma\nu}\right],$$ where the metric gets small perturbations of the kind: $g_{\mu\nu}=g^{(0)}_{\mu\nu}+h_{\mu\nu}$ (here $g^{(0)}_{\mu\nu}$ is the background metric while $h_{\mu\nu}\ll 1$ are the small perturbations), yields to the gravitational spectrum consisting of:\footnote{For the full details of the linearization procedure see \cite{bueno-prd-2016}.} i) a massless graviton, ii) a massive (ghost) graviton with mass $m_g$ and a massive scalar mode with mass $m_s$, i. e., basically the same spectrum found in the four-dimensional quadratic theory of \cite{stelle, stelle-1, hindawi, hindawi-1}. In the effective theory, in the limit $|m_g|\rightarrow\infty$, $|m_s|\rightarrow\infty$, the massive vacuum modes become infinitely heavy and decouple from the spectrum of the theory, leaving the massless graviton as the only propagating vacuum degree of freedom, as in GR. It is demonstrated in \cite{bueno-prd-2016} that the most general cubic theory possessing dimension-independent couplings, which shares spectrum with GR reads: 

\bea {\cal L}=\frac{\sqrt{|g|}}{2}\left(R-2\Lambda_0+2\alpha{\cal X}_4+2\beta{\cal X}_6+2\lambda{\cal P}\right),\label{pablo-lag-fin}\eea where, in four dimensions, the quadratic Lovelock term ${\cal X}_4$ is topological, while the cubic Lovelock term ${\cal X}_6$ identically vanishes. The cubic term ${\cal P}$:

\bea &&{\cal P}=12R_{\mu\;\;\lambda}^{\;\;\nu\;\;\sigma}R_{\nu\;\;\sigma}^{\;\;\tau\;\;\rho}R_{\tau\;\;\rho}^{\;\;\mu\;\;\lambda}+R_{\mu\lambda}^{\;\;\;\;\nu\sigma}R_{\nu\sigma}^{\;\;\;\;\tau\rho}R_{\tau\rho}^{\;\;\;\;\mu\lambda}\nonumber\\
&&\;\;\;\;\;\;\;-12R_{\mu\lambda\nu\sigma}R^{\mu\nu}R^{\lambda\sigma}+8R_\mu^{\;\;\lambda}R_\lambda^{\;\;\nu}R_\nu^{\;\;\mu},\label{cubic-pinv}\eea is neither trivial nor topological in four dimensions.

%--------------------------CECG and geom infl----------------

In \cite{arciniega-plb-2020} a cubic modification of Einstein's GR was proposed which generalizes the ECG \cite{bueno-prd-2016, bueno-prd-2016-1, chinese}. The proposed modification, called as cosmological ECG (CECG), rests on the following combination of cubic invariants: ${\cal P}-8{\cal C}$, where

\bea &&{\cal C}=R_{\mu\lambda\nu\sigma}R^{\mu\lambda\nu}_{\;\;\;\;\;\;\tau}R^{\sigma\tau}-\frac{1}{4}\,RR_{\mu\lambda\nu\sigma}R^{\mu\lambda\nu\sigma}\nonumber\\
&&\;\;\;\;\;\;\;-2R_{\mu\lambda\nu\sigma}R^{\mu\nu}R^{\lambda\sigma}+\frac{1}{2}\,RR_{\mu\nu}R^{\mu\nu}.\label{cubic-inv}\eea Although this latter invariant was previously found in \cite{mann-prd-2017}, in that reference the authors were interested in static spherically symmetric spaces where ${\cal C}$ vanishes. The action of the CECG theory and the derived equations of motion read (we adopt $\mathcal{R}_{(3)}\equiv\mathcal{P}-8\mathcal{C}$):

\bea &&S=\frac{1}{2}\int d^4x\sqrt{|g|}\left(R-2\Lambda+2\beta\mathcal{R}_{(3)}\right),\nonumber\\
&&2{\cal E}_{\mu\nu}=G_{\mu\nu}+g_{\mu\nu} \Lambda+2\beta\left[\left(\frac{\partial\mathcal{R}_{(3)}}{\partial R^{\mu\alpha\beta\sigma}}\right)R_{\nu}^{\,\,\,\alpha\beta\sigma}\right.\nonumber\\
&&\;\;\;\;\;\;\;\;\;\left.-\frac{1}{2}g_{\mu\nu}\mathcal{R}_{(3)}-2\nabla^\alpha\nabla^\beta\left(\frac{\partial \mathcal{R}_{(3)}}{\partial R^{\mu\alpha \beta \nu}}\right)\right]=0.\label{moteq-fin}\eea In general the equations \eqref{moteq-fin} are fourth-order and so the Lovelock theorem \cite{lovelock} is not violated by the CECG theory. An interesting property of this theory is that in Friedmann-Robertson-Walker (FRW) spacetime the motion equations are second order in the time-derivatives. For other backgrounds as, for instance, the plane-symmetric Bianchi I space, the cosmological equations are fourth-order in the derivatives (see the appendix A of \cite{staro-arxiv}).

In \cite{oliva} it is shown that the combination of cubic invariants defining five-dimensional quasitopological gravity, when written in four dimensions, reduce to the CECG. It is also introduced a quartic version of the CECG and a combination of quintic invariants with the properties of the mentioned theory. In this context we want to mention that long ago, in \cite{bento}, the effect of higher-curvature terms, up to corrections quartic in the curvature invariants, in the string low-energy effective action has been studied for the bosonic and heterotic strings as well as for type II superstring in the cosmological context.\footnote{In the context of superstring theory it has been shown long ago that the effective gravitational action should be, at least, 4th order in the Riemann tensor \cite{gross-npb-1986}.} Meanwhile in \cite{arciniega-2} it is shown how to construct invariants up to 8th order in the curvature. In the latter reference it was shown also that the presence of an inflationary epoch is a natural, almost unavoidable, consequence of the existence of a sensible formalism involving an infinite tower of higher-curvature corrections to the Einstein-Hilbert action. The formalism was called ``geometric inflation'' because the only field required was the metric. In string theory
we are familiar with such a structure as the string effective action contains an infinite series of higher curvature corrections to the leading Einstein gravity (see, for instance, \cite{bento}).

%---------------------------dyn syst----------------

The beyond Lovelock theories -- as any other higher curvature modification of GR -- are characterized by the high complexity of their mathematical structure, so that only through feasible approximations one may retrieve some useful analytic information on the cosmological dynamics. Otherwise one has to perform either a numeric investigation or to apply the tools of the dynamical systems theory. The latter allows one to retrieve very useful information on the asymptotic dynamics of the mentioned cosmological models. The asymptotic dynamics may be characterized by either i) attractor solutions to which the system evolves for a wide range of initial conditions, ii) saddle equilibrium configurations that attract the phase space orbits in one direction but repel them in another direction, iii) source critical points which may be pictured as past attractors, or iv) limit cicles, among others. Although the use of the dynamical systems is specially useful when one deals with scalar-field cosmological models (for a small but representative sample see \cite{ellis-book, coley-book, wands-prd-1998, faraoni-grg-2013, bohmer-rev, quiros-rev, quiros-ejp-rev}), its usefulness in other contexts has been explored as well \cite{quiros1, quiros2, quiros3, quiros4}. The tools of the dynamical systems have been used, in particular, in the study of the CECG cosmological model in \cite{quiros-plb-sbmtd}, while in \cite{marciu} the dynamics of the so called extended cubic gravity $f({\cal P})$ was explored.

%-------------------------aim of the paper---------------------

In the present paper we shall look for the global asymptotic dynamics of the geometric inflation formalism developed in \cite{arciniega-2}. We shall explore two explicit toy models where it is possible to sum the infinite series of higher curvature corrections that arises in the formalism. These would be very interesting possibilities if regard gravity as a quantum effective theory \cite{donoghue}. Actually, as discussed in the latter reference, if treat gravity as a quantum effective theory, it is a well-behaved quantum theory at low energies. For the gravitational part of the effective theory we would have:

\bea &&S^\text{eff}_\text{grav}=\int d^4x\sqrt{|g|}\left(\frac{1}{2}R-\Lambda\right.\nonumber\\
&&\;\;\;\;\;\;\;\;\;\;\;\;\;\;\;\;\;\;\;\;\;\;\;\;\;\;\;\left.+c_1R^2+c_2R_{\mu\nu}R^{\mu\nu}+\cdots\right),\label{eff-grav}\eea where $\Lambda$, $c_1$, $c_2$,... are constants and the ellipses denote higher powers of $R$, $R_{\mu\nu}$ and $R_{\mu\nu\sigma\lambda}$. At one loop the divergences due to the massless gravitons read \cite{veltmann} (recall that we work with units where $8\pi G_N=1$): $$\Delta{\cal L}^{(1)}=\frac{1}{8\pi^2\epsilon}\left(\frac{1}{120}R^2+\frac{7}{20}R_{\mu\nu}R^{\mu\nu}\right),$$ where the constant $\epsilon=4-d$ within dimensional regularization. At two loops these divergences have the form \cite{goroff}: $$\Delta{\cal L}^{(2)}=\frac{209}{1440(16\pi^2)^2\epsilon}R^{\alpha\beta}_{\;\;\;\mu\nu}R^{\mu\nu}_{\;\;\;\sigma\lambda}R^{\sigma\lambda}_{\;\;\;\alpha\beta}.$$ As properly noted in \cite{donoghue}, the key feature is that higher powers of $R$, $R_{\mu\nu}$ and $R_{\mu\nu\sigma\lambda}$, are involved at higher loops. In this regard the two toy models proposed in \cite{arciniega-2} would represent an interesting possibility to consider all of the higher order modifications of GR, i. e., all of the higher order graviton loops, without involving a perturbative approach. One would naively expect that, consideration of the whole infinite tower of curvature invariants, would amount to consideration of all of the higher order loops. Hence, quantum effects would be manifest, at least, at high curvature regime.

Our aim is to corroborate, from the dynamical systems perspective, the result of \cite{arciniega-2} that primordial inflation is a generic outcome of the resulting cosmological model. While doing so, the role of the new scale $L\gtrsim L_\text{Pl}$ ($L_\text{Pl}$ is the Planck length), will be revealed as well. We will be able to connect the inflationary stage with effects that arise at curvature scales $\sim L^{-2}_\text{Pl}\gtrsim L^{-2}$, so that these are necessarily quantum effects. We shall show that, in the ``classic limit'', i. e., in the limit when the coupling of the higher curvature corrections vanishes, the primordial inflation of quantum origin is replaced by a bigbang singularity.

%---------------------------organization----------------------

We have organized the paper in the following way. In section \ref{sect-setup} the basic elements of the geometric inflation formalism are given. In section \ref{sect-din-sist} we expose the main properties of the dynamical system corresponding to the two toy models proposed in \cite{arciniega-2}, where the sum of the infinite tower of higher curvature corrections to gravity is explicitly computed. The global dynamics of the mentioned toy models are discussed in section \ref{sect-g-dyn} where the results of the dynamical systems study are presented and physically analyzed. In section \ref{sect-discuss} we discuss on the physical consequences of the obtained results, while brief conclusions are given in section \ref{sect-conclu}. Unless otherwise stated, here we use the units where $8\pi G_N=M^{-2}_\text{Pl}=c^2=1$ ($M_\text{Pl}$ is the Planck mass).

\bigskip

%%%%%%%%%%%%%%%%%%%%%%%%%%%%%%%%%%%%%%%%%%%%%%%%%%%%%%%%%%%%%%%%%%%%%%%%%%%%%%%%%%%%

\section{The basics of the geometric inflation formalism}\label{sect-setup}

Here we consider the formalism proposed in \cite{arciniega-2} that is given by the following action:

\bea S=\frac{1}{2}\int d^4x\sqrt{|g|}\left[R-2\Lambda+\sum_{n=3}^\infty\lambda_n L^{2n-2}{\cal R}_{(n)}\right],\label{action}\eea where ${\cal R}_{(n)}$ are densities constructed from contractions of the metric and the Riemann tensor, $\lambda_n$ are dimensionless constants and $L^{-1}$ is a new energy scale below the Planck scale, $L^{-1}\lesssim L^{-1}_\text{Pl}$. For $L^{-1}\ll L^{-1}_\text{Pl}$ (and $\lambda_3\neq 0$) the theory is affected by causality issues due to the presence of an infinite tower of massive higher spin particles \cite{maldacena}. Besides, as stated in \cite{arciniega-2}, the most reasonable choice for the new scale $L^{-1}$ seems to be that it corresponds to some new scale below the Planck mass, but high enough to make the higher-curvature effects become negligible at late times. Here, in line with the former analysis, we shall assume that $L^{-1}\lesssim L^{-1}_\text{Pl}$.

Among other desirable properties, the geometric inflation theory is ghost-free around maximally symmetric backgrounds and the FRW cosmological equations of motion are second order such as these are for the CECG theory \cite{arciniega-plb-2020}. Actually, in terms of the FRW line-element with flat spatial sections: 

\bea ds^2=-dt^2+a^2(t)\delta_{ik}dx^idx^k,\nonumber\eea the cosmological equations of motion derived from \eqref{action} with the addition of a matter piece of action read \cite{arciniega-2}:

\bea &&3F(H)=\rho_m+\Lambda,\nonumber\\
&&\frac{\dot H}{H}F'(H)=-(p_m+\rho_m),\nonumber\\
&&\dot\rho_m=-3H(\rho_m+p_m),\label{moteq}\eea where $H$ is the Hubble parameter, $\rho_m$ and $p_m$ are the energy density and pressure of the matter fluid\footnote{In what follows, for simplicity, we assume the following equation of state (EOS) for the matter fluid: $p_m=\omega_m\rho_m$, where the constant $\omega_m$ is the EOS parameter.} and $F'\equiv dF/dH$. The function $F=F(H)$ reads:

\bea F(H)=H^2+L^{-2}\sum_{n=3}^\infty (-1)^n\lambda_n\left(LH\right)^{2n}.\label{fh}\eea 

In \cite{arciniega-2} two different kinds of conditions were given on the dimensionless parameters $\lambda_n$ such that the infinite summation in \eqref{fh} can be explicitly performed.\footnote{As clearly stated in \cite{arciniega-2} many other summable choices are possible.} These conditions led to two different toy models that are based in the following forms of the function $F$:

\bea F(H)=H^2\left[1+\lambda_4\left(LH\right)^6e^{\left(LH\right)^4}\right],\label{mod1}\eea for model 1 and 

\bea F(H)=H^2\left\{1-\lambda_3\left(LH\right)^4\left[1-\left(LH\right)^2e^{\left(LH\right)^2}\right]\right\},\label{mod2}\eea for model 2. Since in these models it is possible to sum over the infinite tower of higher-order curvature contributions, we expect that at high curvature $H^2\gtrsim L^{-2}$ quantum gravitational effects would become important, if the present classical theory is regarded as an effective quantum theory \cite{donoghue}. As a matter of fact, the Planck length represents the boundary of the quantum domain, so that below $L_\text{Pl}$, i. e., at very high curvature $H^2\gtrsim L^{-2}_\text{Pl}$, quantum gravity is the dominating contribution. Strictly speaking, we would trust the present formalism up to curvatures $\sim L^{-2}_\text{Pl}$.

%%%%%%%%%%%%%%%%%%%%%%%%%%%%%%%%%%%%%%%%%%%%%%%%%%%%%%%%%%%

\section{The Dynamical System}\label{sect-din-sist}

Here we follow quite a different approach than in \cite{quiros-plb-sbmtd}. We choose variables of some phase space that are dimensionless and bounded as in the mentioned reference, but the constants of the theory $\lambda_k$, where $k=4$ for model 1, while $k=3$ for model 2, are not absorbed into these variables. Instead these remain as free constants of the dynamical system, affecting the existence and stability of the equilibrium configurations.

%=========================================

\subsection{Model 1}\label{subsect-mod1}

The cosmological equations of motion \eqref{moteq} for the choice \eqref{mod1} read:

\bea &&1+\lambda_4L^6H^6e^{L^4H^4}=\Omega_m+\Omega_\Lambda,\nonumber\\
&&-2\frac{\dot H}{H^2}=\frac{3(\omega_m+1)\Omega_m}{1+2\lambda_4L^6H^6e^{L^4H^4}\left(2+L^4H^4\right)},\label{moteq-m1}\eea where, as customary, $\Omega_m\equiv\rho_m/3H^2$ is the dimensionless energy density of the matter degrees of freedom, while $\Omega_\Lambda=\Lambda/3H^2$.

Here, in order to investigate the global asymptotic dynamics of this model, we introduce the following bounded variables of some phase space:

\bea &&x\equiv\frac{1}{1+L^2H^2}\;\Rightarrow\;L^2H^2=\frac{1-x}{x},\nonumber\\
&&y\equiv\frac{1}{1+\Omega_m}\;\Rightarrow\;\Omega_m=\frac{1-y}{y},\label{vars-1}\eea where $0\leq x\leq 1$ and $0\leq y\leq 1$. The modified Friedmann constraint -- first equation in \eqref{moteq-m1} -- can be written in the following way:

\bea \Omega_\Lambda=\frac{2y-1}{y}+\frac{\lambda_4(1-x)^3e^{\left(\frac{1-x}{x}\right)^2}}{x^3},\label{fried-const-1}\eea meanwhile,

\bea \frac{\dot H}{H^2}=-\frac{3(\omega_m+1)x^2A(x)(1-y)}{2B(x)y},\label{hdot-1}\eea where, for compactness of writing, we have introduced the functions:

\bea &&A(x)=x^3e^{-\left(\frac{1-x}{x}\right)^2},\nonumber\\
&&B(x)=x^5e^{-\left(\frac{1-x}{x}\right)^2}\nonumber\\
&&\;\;\;\;\;\;\;\;\;\;\;\;+2\lambda_4(1-x)^3(3x^2-2x+1),\label{ab}\eea respectively. 

In terms of the phase space variables $x$, $y$, the second-order cosmological equations \eqref{moteq} may be traded by the following two-dimensional autonomous dynamical system:

\bea &&\frac{dx}{dv}=\frac{3(\omega_m+1)x^3(1-x)A(x)(1-y)}{B(x)},\nonumber\\
&&\frac{dy}{dv}=3(\omega_m+1)y(1-y)\left[y-\frac{x^2A(x)(1-y)}{B(x)}\right],\label{ode-1}\eea where we have introduced the time variable 

\bea v=\int(1+\Omega_m)Hdt.\label{v}\eea 

The phase space where to look for equilibrium configurations of the dynamical system \eqref{ode-1} is the following unit phase square:

\bea \Psi=\left\{\left(x,y\right):\,0\leq x\leq 1,\;0\leq y\leq 1\right\}.\label{psi}\eea The separatrix

\bea {\cal S}=\left\{\left(x,y\right):\,0\leq x\leq 1,\,y=\bar y_1(x)\right\},\label{sep}\eea where

\bea \bar y_1(x)=\frac{A(x)}{2A(x)+\lambda_4(1-x)^3},\label{sep-y}\eea separates the region where the backgound space is de Sitter, $\Omega_\Lambda\geq 0$ $\Rightarrow y\geq\bar y_1(x)$, from the region where the background space is anti-de Sitter, $\Omega_\Lambda<0$ $\Rightarrow y<\bar y_1(x)$. Here we concentrate in de Sitter background spaces exclusively, so that we shall consider only the region of the phase square above the separatrix: 

\bea \Psi_{\text{phys},1}=\left\{\left(x,y\right):\,0\leq x\leq 1,\;\bar y_1(x)\leq y\leq 1\right\}.\label{psi-phys}\eea

Another curve of physical interest is the one related with the change of sign of the deceleration parameter:

\bea q\equiv-1-\frac{\dot H}{H^2},\label{q}\eea i. e., the curve that follows from the condition $q=0$,

\bea \hat y_1(x)=\frac{3(\omega_m+1)x^2A(x)}{3(\omega_m+1)x^2A(x)+2B(x)}.\label{q0}\eea Accelerated expansion occurs whenever $y>\hat y_1(x)$.

%=========================================

\subsection{Model 2}\label{subsect-mod2}

Here, as in the former subsection, we shall focus in de Sitter background spaces exclusively, so that only the case with $\Lambda\geq 0$ will be of interest. The cosmological equations of motion \eqref{moteq} for the choice \eqref{mod2} read:

\bea &&1-\lambda_3L^4H^4+\lambda_3L^6H^6e^{L^2H^2}=\Omega_m+\Omega_\Lambda,\nonumber\\
&&\frac{\dot H}{H^2}=\frac{-3(\omega_m+1)\Omega_m/2}{1-3\lambda_3L^4H^4+\lambda_3\left(4+L^2H^2\right)L^6H^6e^{L^2H^2}},\nonumber\\
&&\dot\Omega_m=-H\Omega_m\left[3(w_m+1)+2\frac{\dot H}{H^2}\right],\label{moteq-m2}\eea where $\lambda_3$ is a dimensionless coupling constant and, as before, $\Omega_m\equiv\rho_m/3H^2$ while $\Omega_\Lambda=\Lambda/3H^2$. 

We shall use the same variables \eqref{vars-1}. We get that,

\bea \Omega_\Lambda=\frac{2y-1}{y}-\lambda_3\left(\frac{1-x}{x}\right)^2\left[1-\left(\frac{1-x}{x}\right)e^\frac{1-x}{x}\right],\label{ol-mod2}\eea and

\bea \frac{\dot H}{H^2}=-\frac{3(\omega_m+1)x^4(1-y)}{2D(x)y},\label{hdot-mod2}\eea where we have introduced the functions:

\bea &&C(x)=3x^2-(1+3x)(1-x)e^\frac{1-x}{x},\nonumber\\
&&D(x)=x^4-\lambda_3(1-x)^2C(x).\label{cd}\eea The following ODE-s are obtained out of \eqref{vars-1} and \eqref{moteq-m2}:

\bea &&x'=-2x(1-x)\frac{\dot H}{H^2},\nonumber\\
&&y'=y(1-y)\left[3(\omega_m+1)+2\frac{\dot H}{H^2}\right],\nonumber\eea where the prime denotes derivative with respect to the time variable $\tau=\ln a$. In a more explicit form the above equations can be written as it follows:

\bea &&\frac{dx}{dv}=\frac{3(w_m+1)x^5(1-x)(1-y)}{D(x)},\nonumber\\
&&\frac{dy}{dv}=3(w_m+1)y(1-y)\left[y-\frac{x^4(1-y)}{D(x)}\right],\label{ode-mod2}\eea where we have used the time variable $v$ in \eqref{v}, instead of $\tau=\int Hdt$. 

Since here we consider de Sitter background spaces exclusively $\Omega_\Lambda\geq 0$, then from \eqref{ol-mod2} it follows that the physically meaningful region of the phase space $\Psi=\{(x,y):0\leq x\leq 1,\,0\leq y\leq 1\}$ is the one located above the separatrix $y=\bar y_2(x)$, where

\bea \bar y_2(x)=\frac{x^3}{2x^3-\lambda_3(1-x)^2\left[x-(1-x)e^\frac{1-x}{x}\right]},\label{sep-mod2}\eea i. e., 

\bea \Psi_{\text{phys},2}=\left\{(x,y):0\leq x\leq 1,\,\bar y_2(x)\leq y\leq 1\right\}.\label{psi-mod2}\eea 

Accelerated expansion occurs for points above the curve $y=\hat y_2(x)$:

\bea \hat y_2(x)=\frac{3(w_m+1)x^4}{3(w_m+1)x^4+2D(x)}.\label{q0-mod2}\eea

%-----------------------------------

\begin{figure*}[t!]
\includegraphics[width=4.5cm]{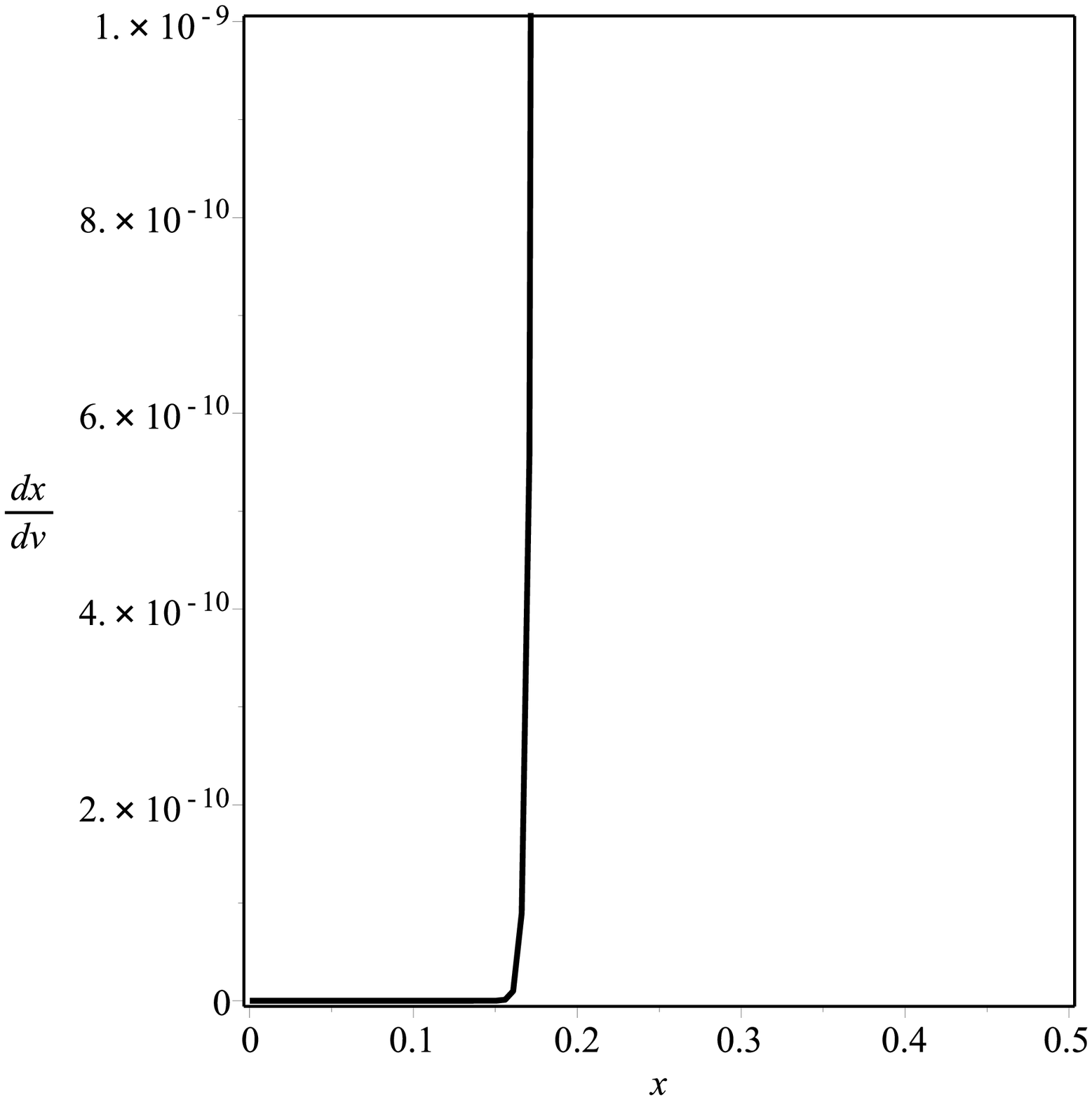}
\includegraphics[width=4.5cm]{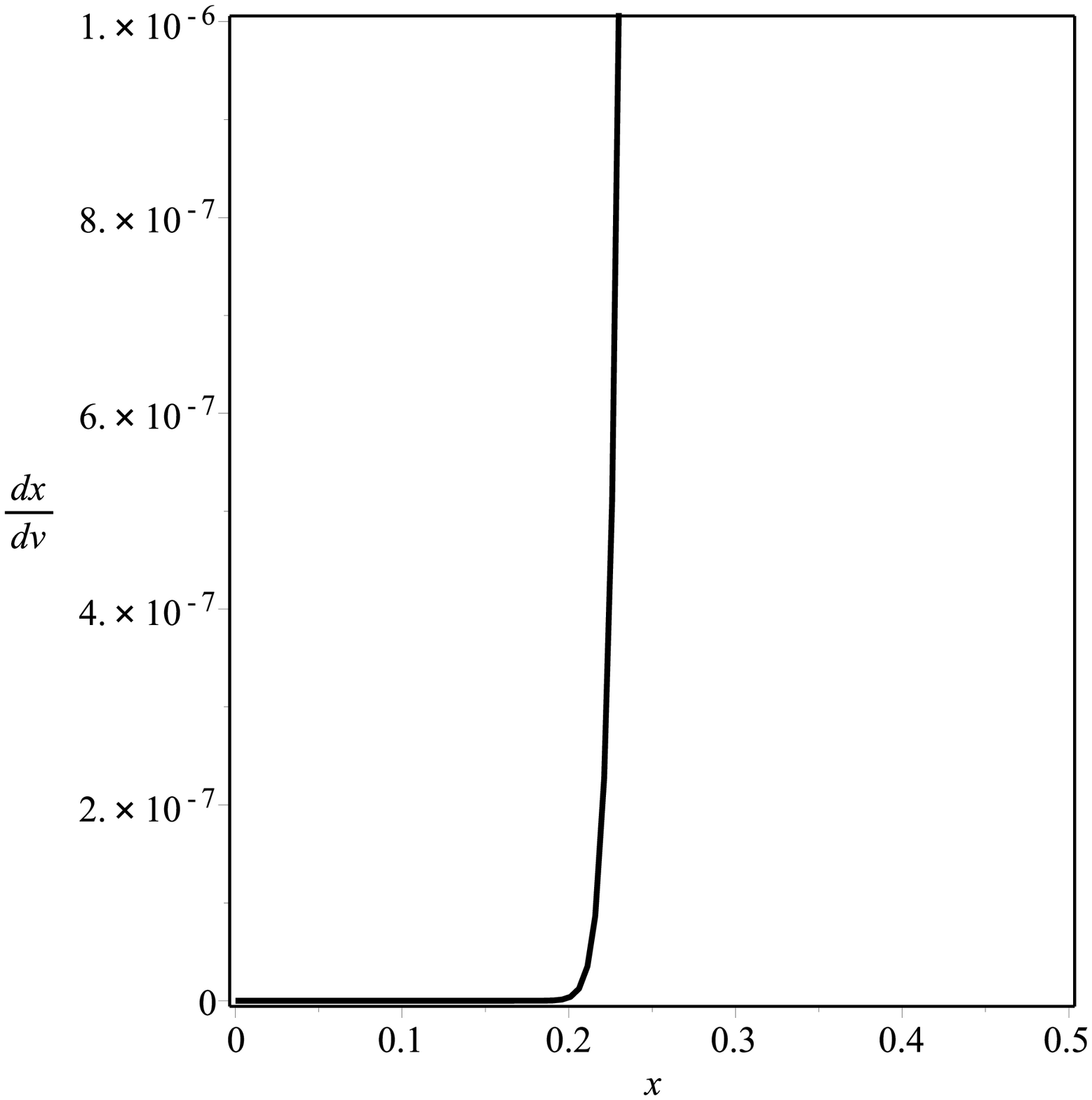}
\includegraphics[width=4.5cm]{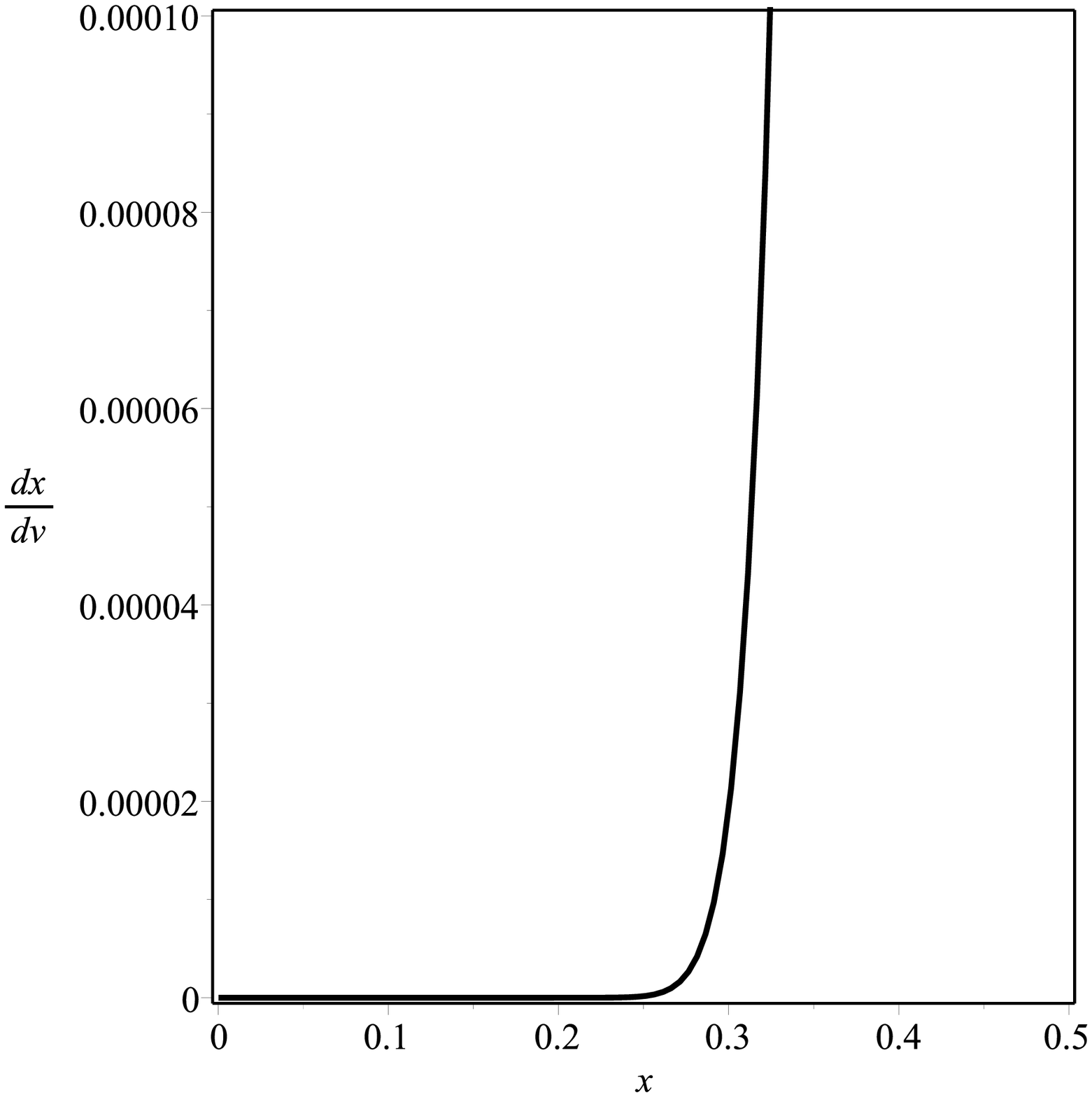}
\includegraphics[width=4.5cm]{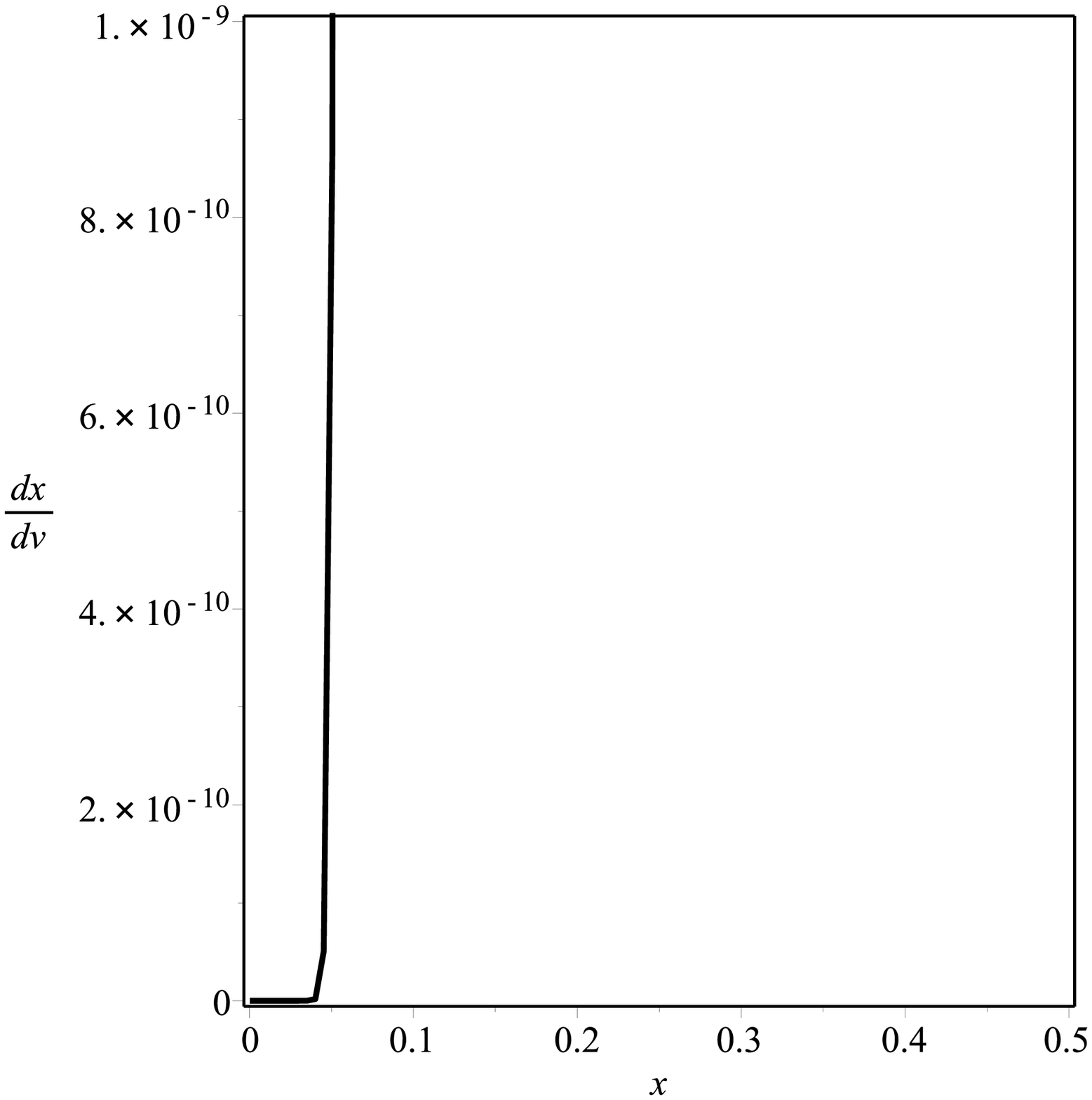}
\includegraphics[width=4.5cm]{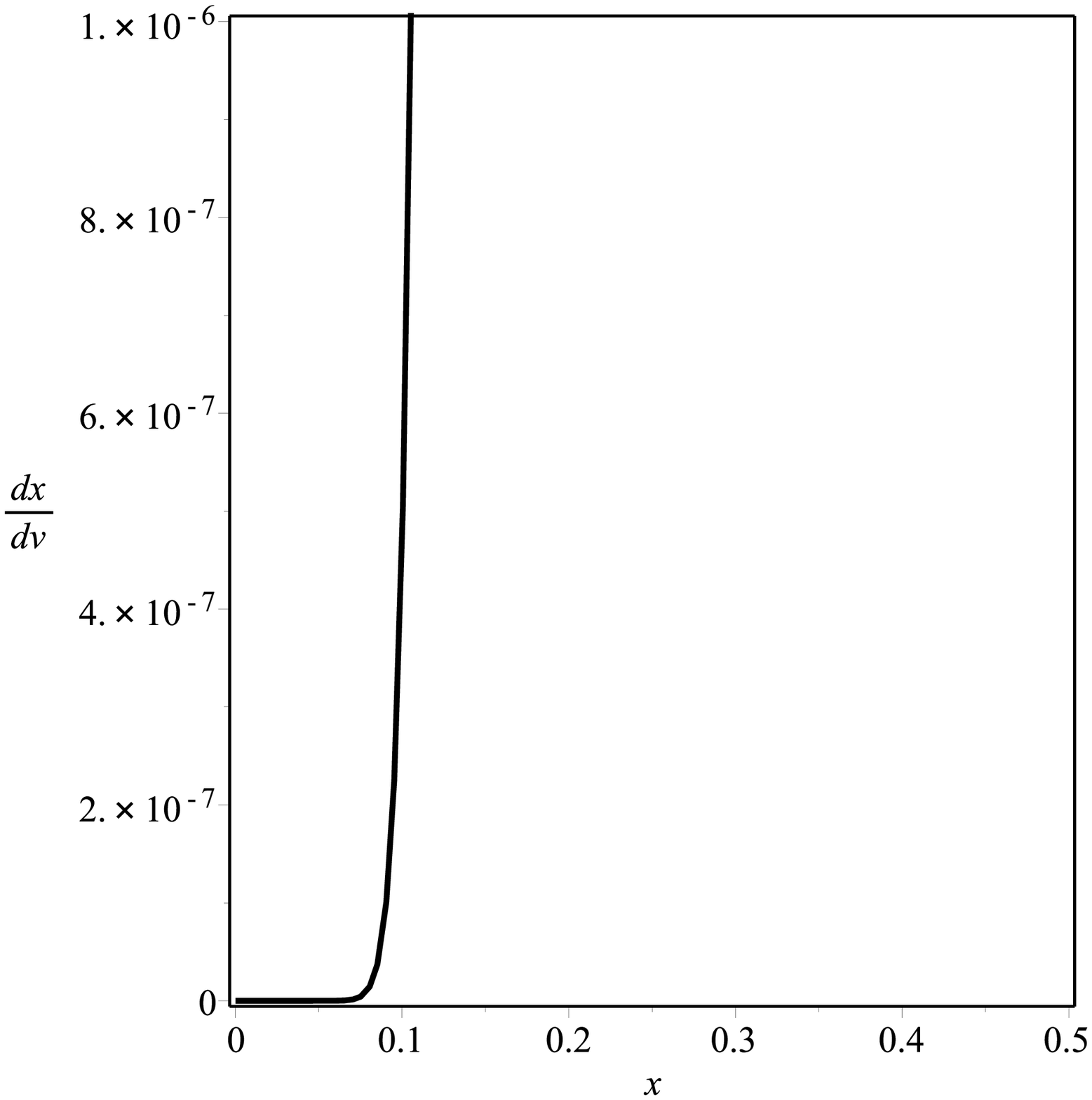}
\includegraphics[width=4.5cm]{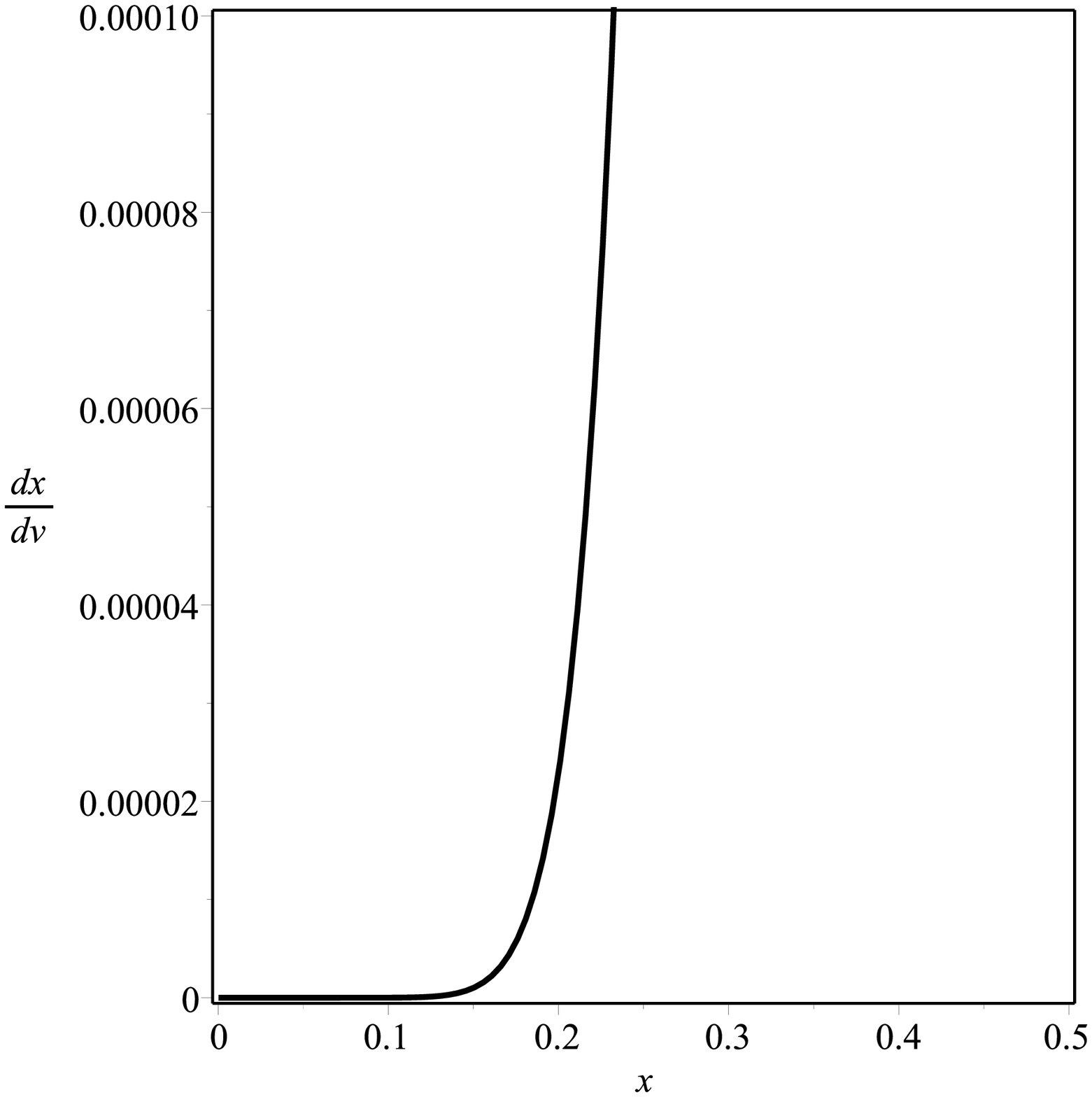}
\vspace{1.2cm}\caption{Plots of $dx/dv$ vs $x$ in the dynamical system for model 1 -- top panels -- and for model 2 -- bottom panels -- for the radiation, for different choices of the dimensionless constants $\lambda_4$ and $\lambda_3$. From left to the right: i) $\lambda_k=10^{-5}$, ii) $\lambda_k=10^{-2}$ and iii) $\lambda_k=1$, where $k=4$ for model 1, while $k=3$ for model 2.}\label{fig0}\end{figure*}

%----------------------------------------

%-----------------------------------

\begin{figure*}[t!]
\includegraphics[width=4cm]{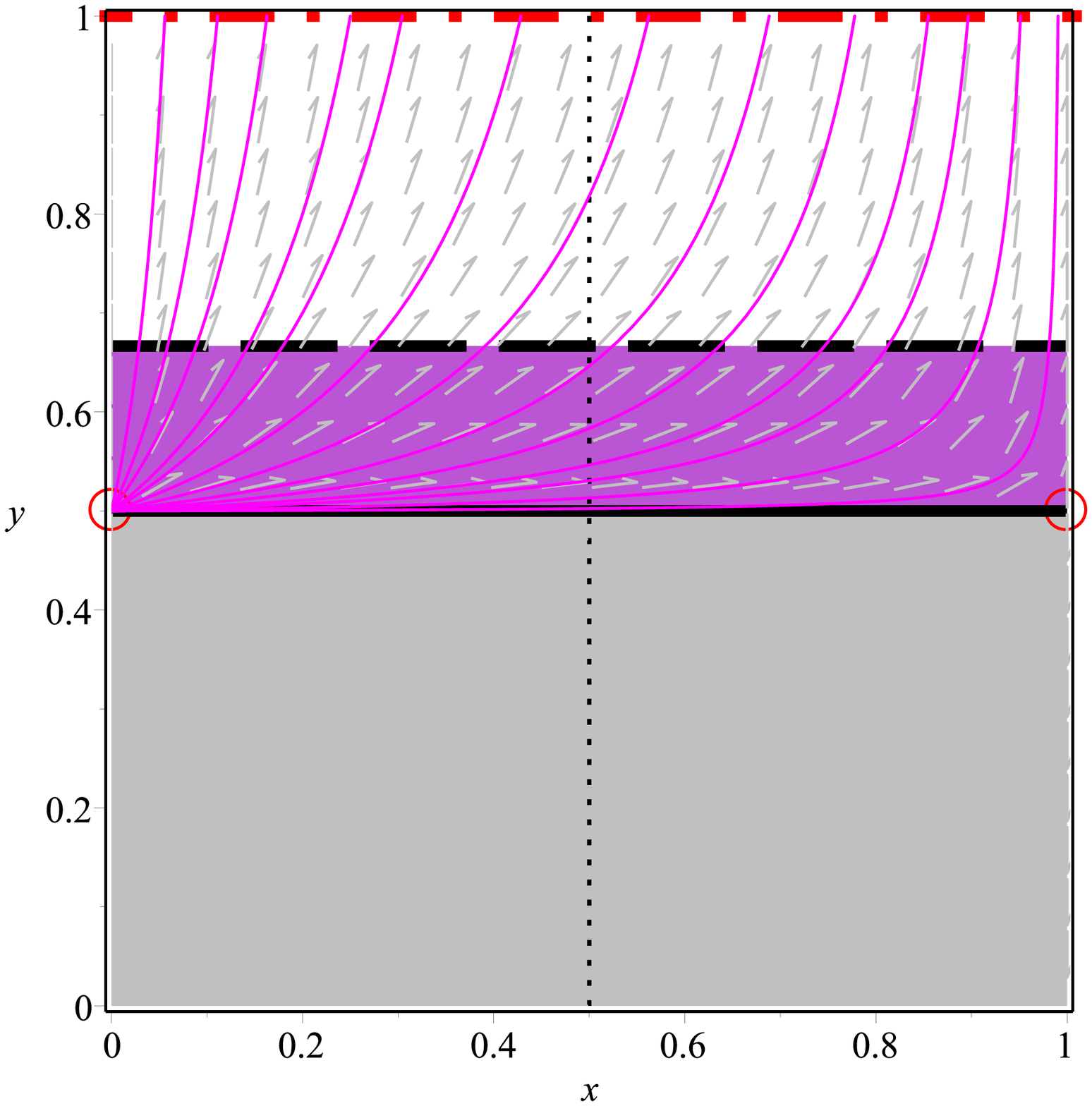}
\includegraphics[width=4cm]{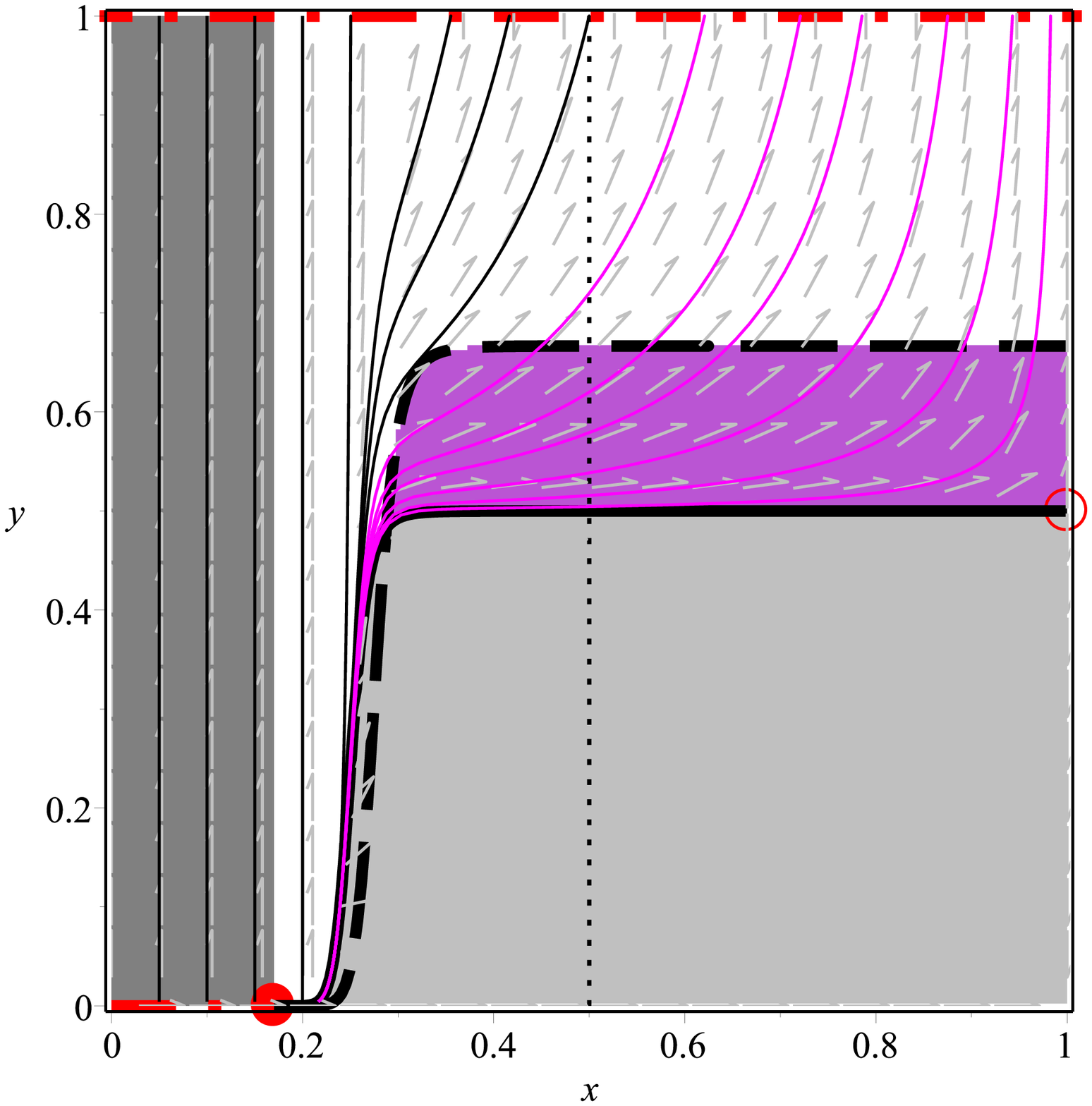}
\includegraphics[width=4cm]{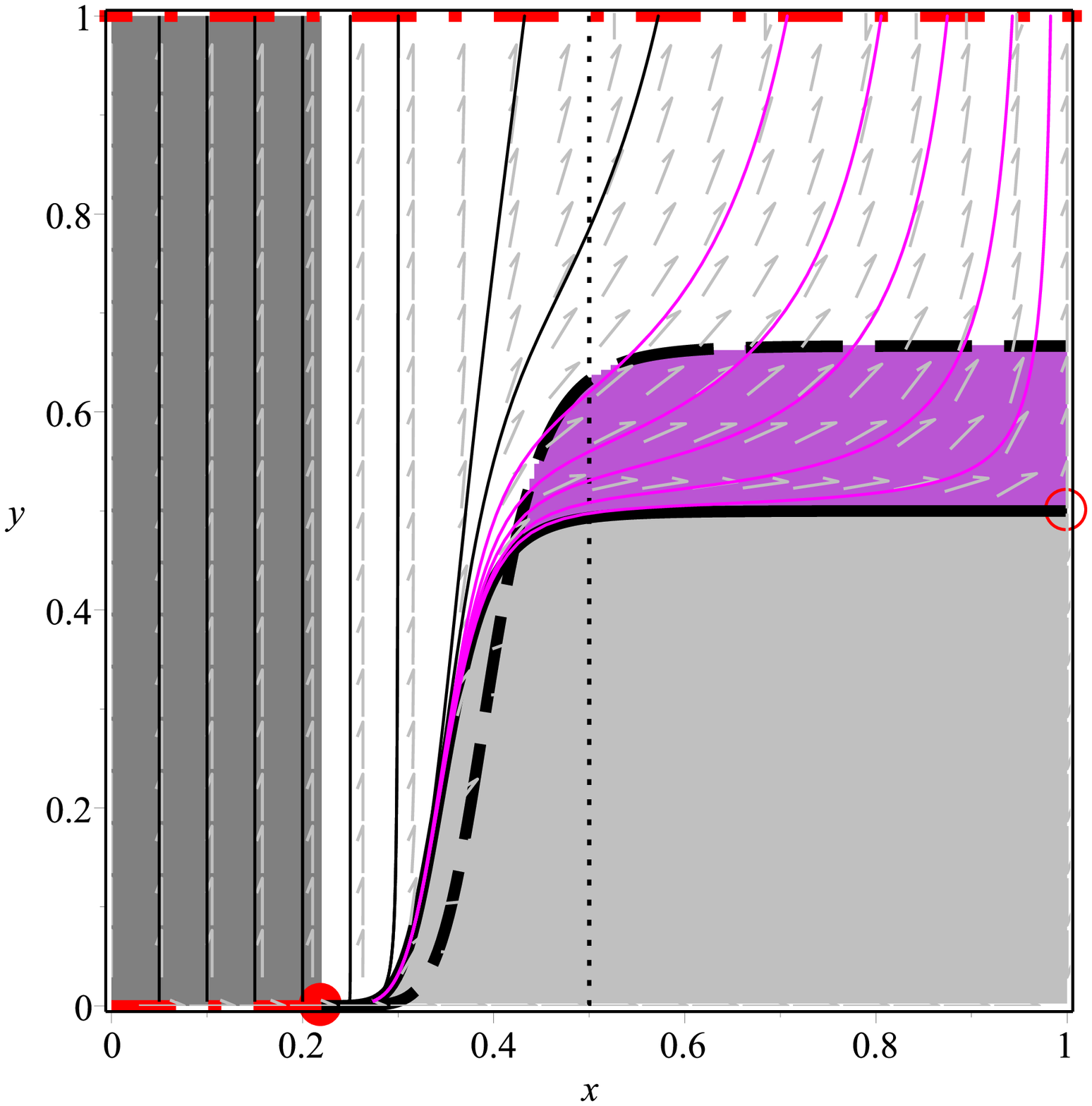}
\includegraphics[width=4cm]{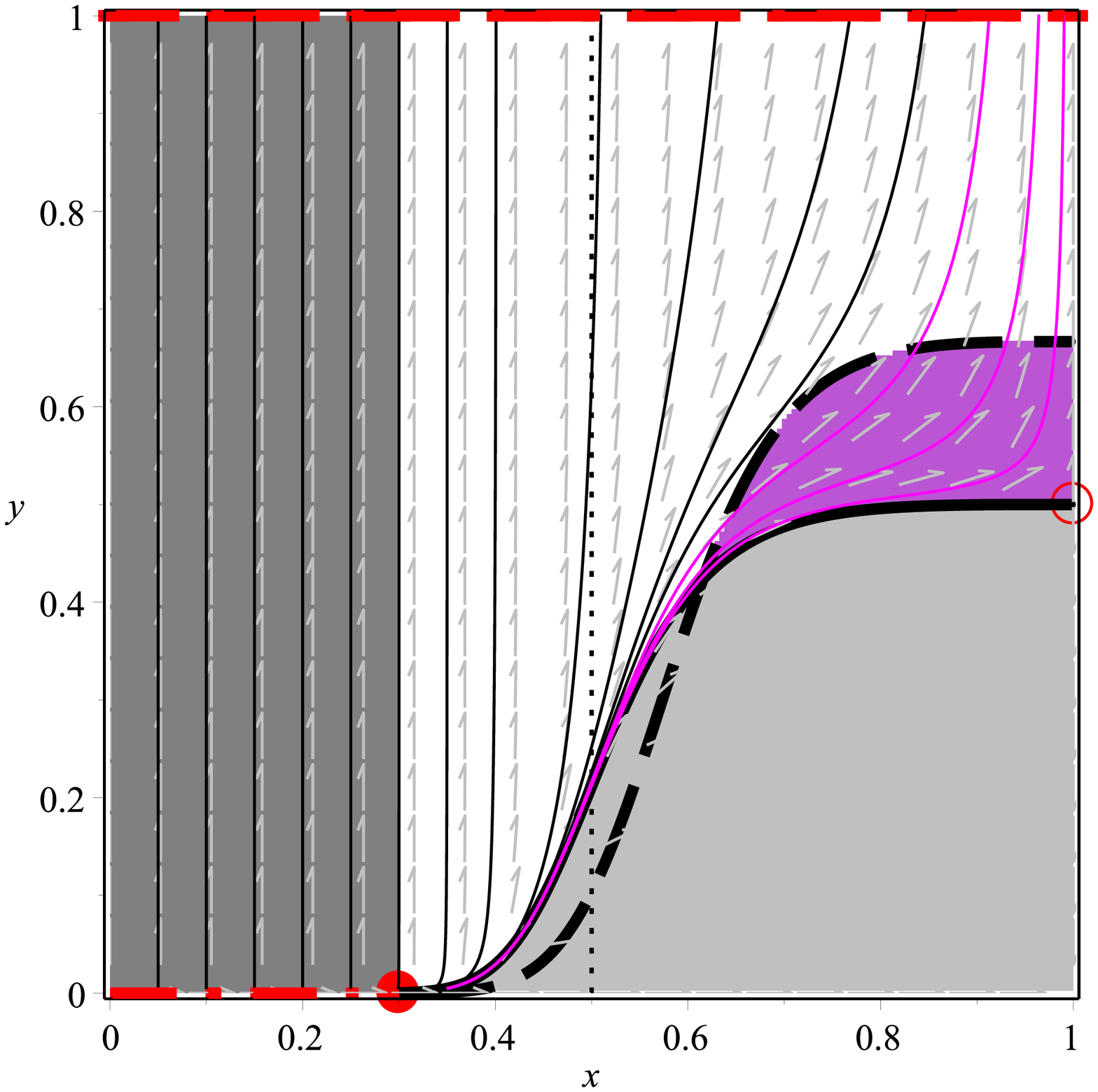}
\includegraphics[width=4cm]{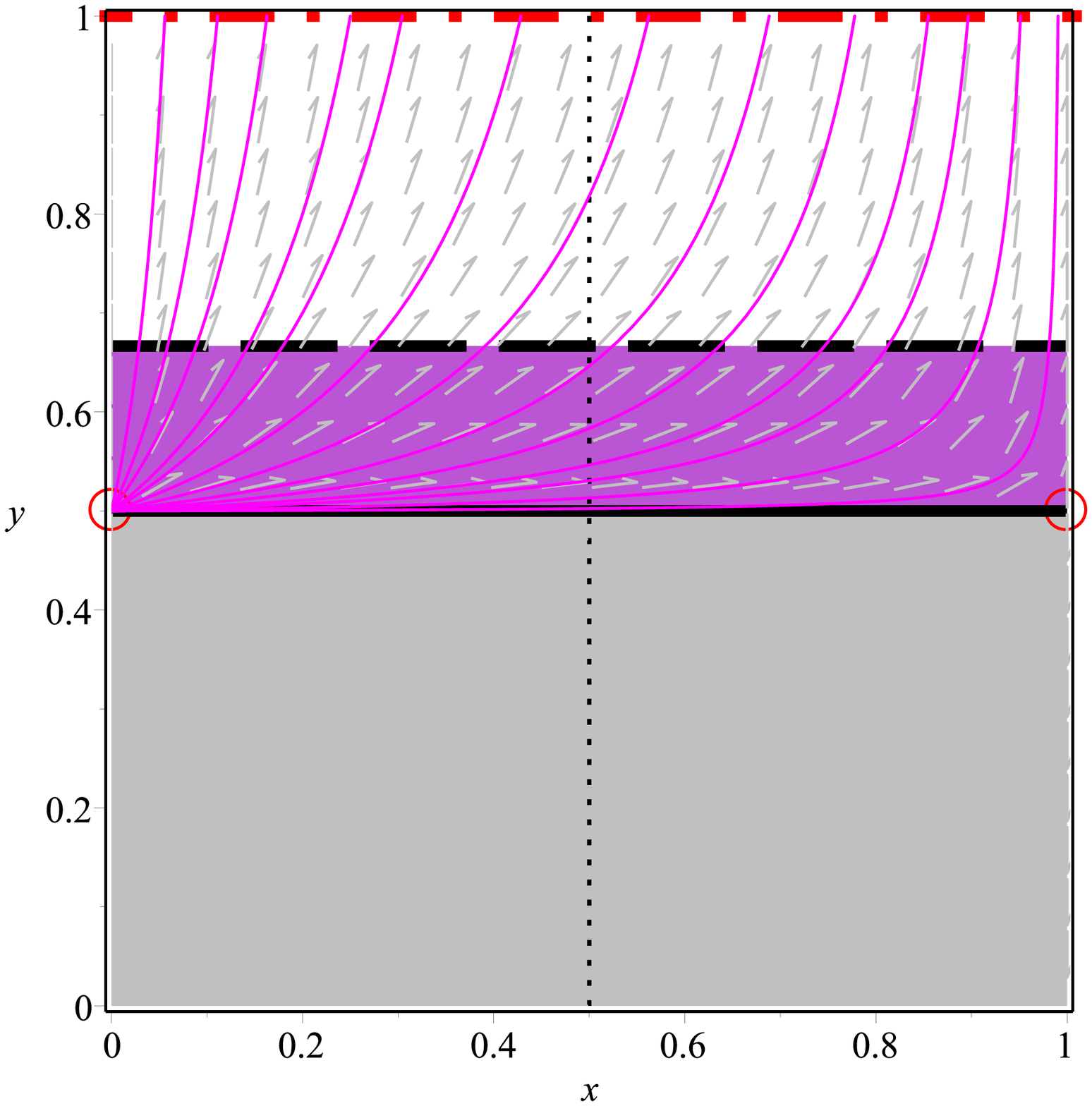}
\includegraphics[width=4cm]{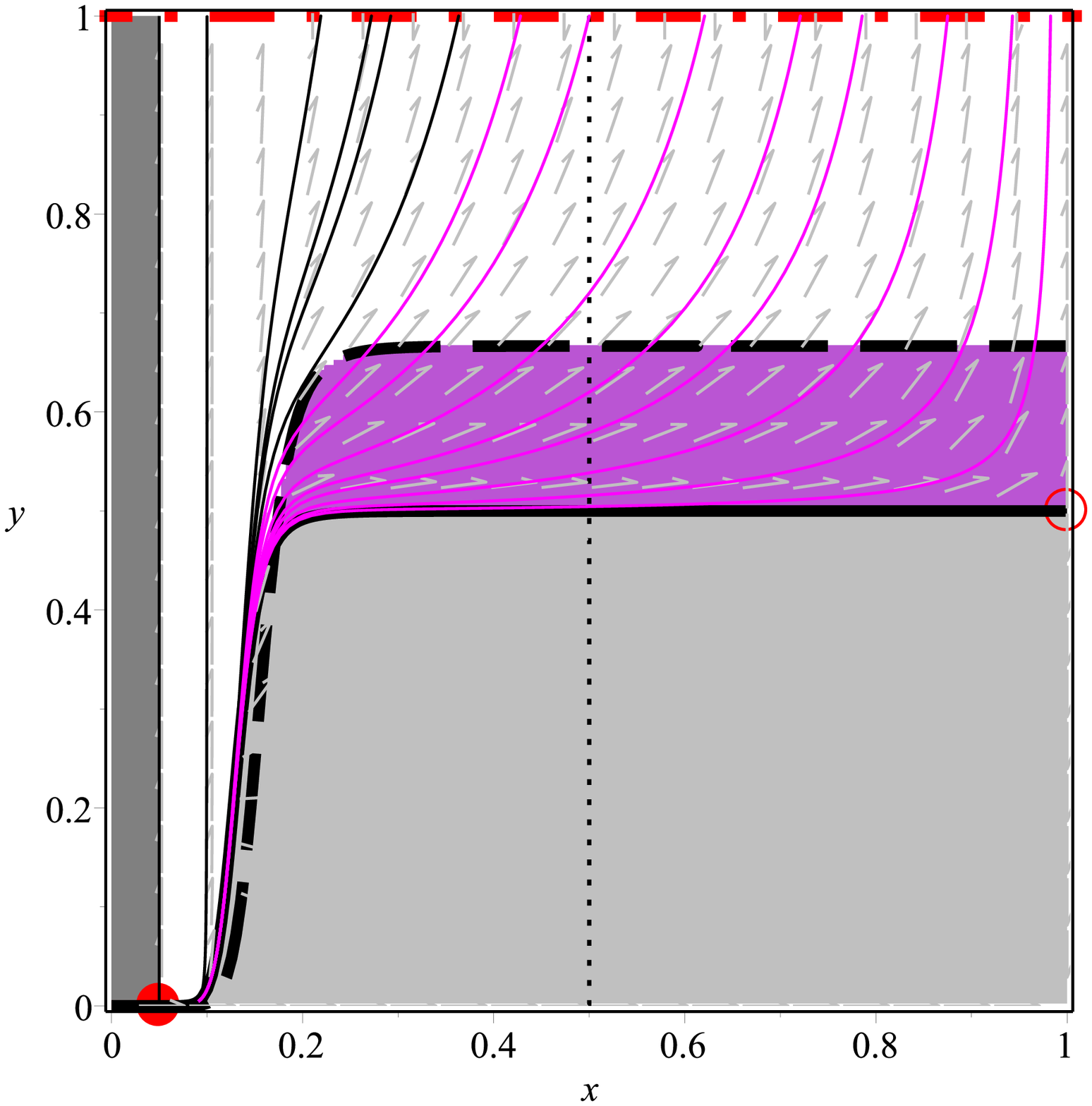}
\includegraphics[width=4cm]{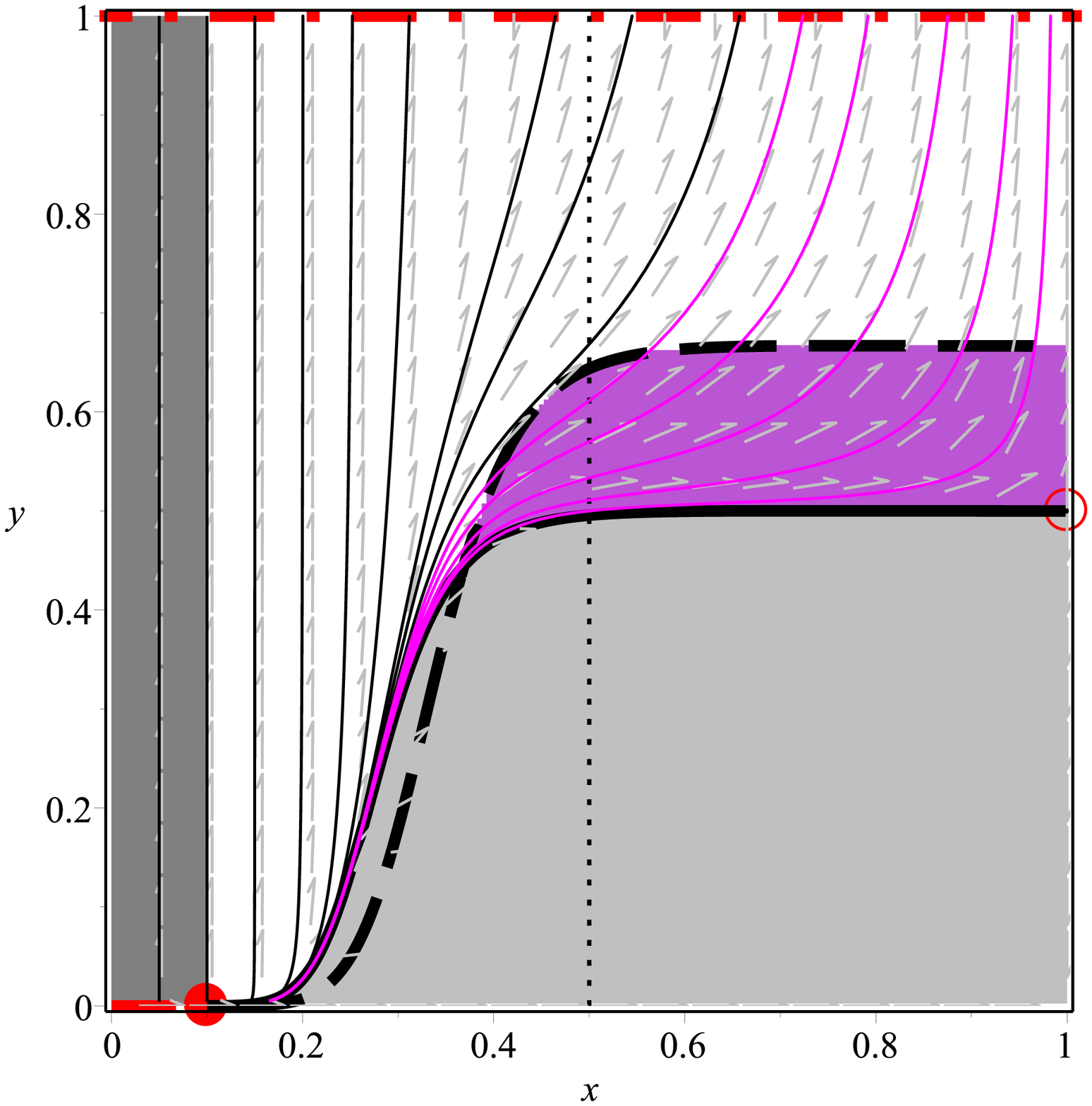}
\includegraphics[width=4cm]{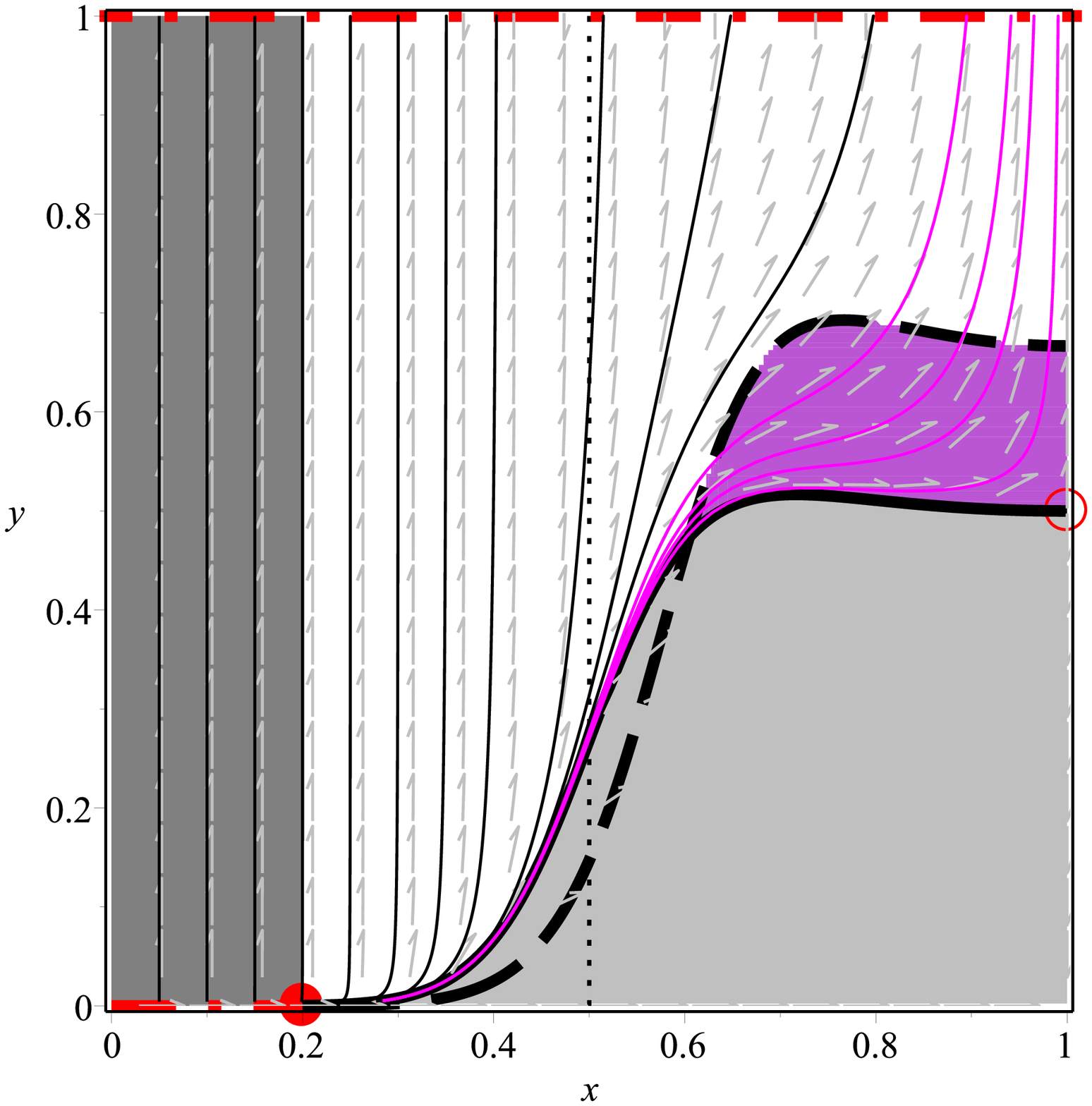}
\vspace{1.2cm}\caption{Phase portrait of the dynamical systems for model 1 -- top panels -- and for model 2 -- bottom panels -- for the radiation ($\omega_m=1/3$) for different choices of the dimensionless constants $\lambda_k$ ($k=4$ for model 1 and $k=3$ for model 2). From left to the right: i) $\lambda_k=0$ ($\Lambda$CDM model), ii) $\lambda_k=10^{-5}$, iii) $\lambda_k=10^{-2}$ and iv) $\lambda_k=1$. The critical points $P_\text{bb}:(0,1/2)$ (exists only for $\lambda_k=0$) and $P_\text{mat}:(1,1/2)$ are enclosed by the small circles, while the de Sitter critical manifold ${\cal M}_\text{dS}$ is represented by the dash-dot line coincident with the upper boundary. The quantum inflationary manifold ${\cal Q}_\text{infl}=\{(x,0):0\leq x\leq\left\langle x\right\rangle\}$ is represented by the dash-dot line at $y=0$ (lower boundary) that starts at the origin and ends up at the solid circle representing the averaged value $\left\langle x\right\rangle$. The vertical dark-gray strip represents the quantum region. Its fuzzy boundary is at $\left\langle x\right\rangle$, i. e., at the mean value of the $x$-coordinate at which $H\approx L^{-1}_\text{Pl}$. The vertical dotted straight line represents the point where $H=L^{-1}$. To the left of this line the energy is greater than the new scale. The separatrix ${\cal S}$ -- solid dark curve -- and the curve corresponding to the condition $q=-1-\dot H/H^2=0$ -- dash dark curve -- are also shown. The region above the separatrix is for de Sitter background ($\Omega_\Lambda>0$) while the region below it (gray shading) corresponds to anti-de Sitter background space which is not of interest here. Above the dashed curve the expansion occurs at an accelerated pace while below it the cosmic expansion is decelerated. The orbits that meet the region with decelerated expansion (magenta shading) are drawn same color as the region.}\label{fig1}\end{figure*}

%----------------------------------------

%%%%%%%%%%%%%%%%%%%%%%%%%%%%%%%%%%%%%%%%%%%%%%%%%%%%%%%%%%%%%

\section{Global dynamics of the models}\label{sect-g-dyn}

Before we list the critical points of the dynamical systems corresponding to the above models and expose their main properties, we want to briefly discuss on an important issue in connection with the existence of a quantum inflationary manifold in the phase space of the models as it is revealed by the numerical investigation. The fact that in the models \eqref{mod1}, \eqref{mod2}, it is possible to sum over the infinite tower of higher order curvature corrections in \eqref{fh}, means that geometric manifestation of quantum effects is possible in these models. If this were so, then these effects should be reflected in the properties of the phase space, including its equilibrium configurations. One would expect, in particular, that the inflationary behavior in the models would be strongly correlated with quantum effects. 

This represents a problem in what regards to the dynamical systems study since, in the quantum domain, due to quantum fluctuations of vacuum, any equilibrium configuration could not strictly satisfy the definition of a critical point, in particular the requirement of exact vanishing of the generalized speed $(dx/dv,dy/dv)=(0,0)$. In this case one should establish a criterion for the accuracy within which to consider vanishing values of the derivatives and of the variables themselves. In the present formalism the dimensionless constants $\lambda_k$ ($k=4$ for model 1 while $k=3$ for model 2) determine the strength of the coupling of higher curvature corrections to gravity. Hence, it seems appropriate to choose these constants as a good standard to establish the mentioned criterion. For purposes of our calculations, here we assume that quantities with values $\leq 10^{-5}\lambda_k$ could be assumed vanishing with an appropriate level of accuracy.

%==========================================================

\subsection{Critical points and their properties}

The critical points $P_i:(x_i,y_i)$ of the dynamical systems \eqref{ode-1} and \eqref{ode-mod2} in the physically meaningful phase spaces $\Psi_{\text{phys},1}$ and $\Psi_{\text{phys},2}$, respectively, as well as their stability properties, are listed and briefly discussed below (see the related FIG. \ref{fig1}).

\begin{enumerate}

\item Quantum inflationary manifold, 

\bea {\cal Q}_\text{infl}=\{(x,0):0\leq x\leq\left\langle x\right\rangle\},\label{binfl}\eea where the length of the manifold is an averaged value. This is a ``fuzzy manifold'' since $\left\langle x\right\rangle$ is not a definite value but, rather, an averaged quantity because, due to the quantum behavior near of the Planck scale $L_\text{Pl}^{-1}$, there are quantum fluctuations. We assume that at the average value $\left\langle x\right\rangle$, $H\approx L^{-1}_\text{Pl}$, so that the boundary of the quantum domain is located, precisely at $\left\langle x\right\rangle$.

This manifold can be found only through numeric investigation, as shown in FIG. \ref{fig0}. Although at $y=0$, the right-hand side of the second equation in the dynamical systems \eqref{ode-1} and \eqref{ode-mod2} exactly vanishes, the right-hand side of the first equation (the one for $dx/dv$) does not vanish exactly but at the origin. Nevertheless, the numerical inspection reveals that there is a region where, within the given accuracy: $dx/dv\lesssim 10^{-5}\lambda_k$, the latter derivative can be assumed vanishing as well (see FIG. \ref{fig0}). This means that it is a critical manifold. 

The fact that all of the relevant quantities evaluated at points in ${\cal Q}_\text{infl}$ blowup: $$\Omega_m\rightarrow\infty,\;\Omega_\Lambda\rightarrow\infty,$$ while $\dot H/H^2\rightarrow 0/0$ is undetermined, is just a manifestation of the quantum nature of points in this manifold. Actually, the quantum inflationary manifold is located inside the quantum domain where the motion equations of the present classical formalism are not supposed to be valid.

From the numerical investigation it follows that critical points in ${\cal Q}_\text{infl}$ are past attractors. In general the manifold is a global past attractor since any possible orbit in the phase space starts at a point in ${\cal Q}_\text{infl}$. Although in the quantum domain we lose track of any classical curve such as, for instance, the separatrix or the curve where $q=0$, immediately after abandoning points in ${\cal Q}_\text{infl}$ the orbits of the phase space enter a region where the expansion is accelerated, this is why we may relate this manifold with inflationary behavior.

\item Standard bigbang solution $P_\text{bb}:(0,1/2)$. In this case $x=0$ $\Rightarrow$ $H\gg L^{-1}$, while $y=1/2$ $\Rightarrow$ $\Omega_m=1$. This solution exists only for $\lambda_k=0$, i. e., in the limit of vanishing coupling of the higher curvature corrections to gravity. Whenever it exists $P_\text{bb}$ is the global past attractor.

\item Matter domination, $P_\text{mat}:(1,1/2)$ $\Rightarrow$ $H\ll L^{-1}$ and $\Omega_m=1$, i. e., $3H^2=\rho_m$. Given that the eigenvalues of the linearization matrix at $P_\text{mat}$: $$\lambda_1=-3(\omega_m+1)/2,\;\lambda_2=3(\omega_m+1)/2,$$ are of different sign, this means that the matter-dominated solution is a saddle critical point. At this solution $\Omega_\Lambda=0$, while $\Omega_m=1$ $\Rightarrow$ $3H^2=\rho_m$, and $$\frac{\dot H}{H^2}=-\frac{3}{2}(\omega_m+1)\Rightarrow q=\frac{3\omega_m+1}{2}.$$

\item de Sitter attractor manifold: 

\bea {\cal M}_\text{dS}=\{(x,1):\;0\leq x\leq 1\},\label{mds}\eea which exists only for de Sitter ($\Lambda>0$) background spaces. For points in ${\cal M}_\text{dS}$ we obtain the following eigenvalues of the corresponding linearization matrix: $\lambda_1=0$, $\lambda_2=-3(w_m+1)$. The vanishing eigenvalue is associated with an eigenvector that is tangent to the manifold at each point. The second eigenvalue is always a negative quantity. This means that, as seen from the FIG. \ref{fig1}, each one of the critical points in ${\cal M}_\text{dS}$ is a local attractor, i. e., the manifold itself is a global attractor of orbits in $\Psi$. For each point in the de Sitter attractor manifold, $\dot H=0$, $\Omega_m=0$ $\Rightarrow q=-1$. Besides: 

\bea \Omega_\Lambda=1+\lambda_4\left(\frac{1-x}{x}\right)^3e^{\left(\frac{1-x}{x}\right)^2},\label{ol-mod11}\eea for model 1, while for model 2: 

\bea \Omega_\Lambda=1-\lambda_3\left(\frac{1-x}{x}\right)^2\left[1-\left(\frac{1-x}{x}\right)e^\frac{1-x}{x}\right],\label{ol-mod21}\eea with $x\in{\cal M}_\text{dS}$ in both cases.

\end{enumerate} 

The above critical points are common to both models 1 and 2. Besides, there are not other equilibrium states in the phase spaces of these models.

%===============================================================================

\subsection{Physical analysis of the phase portrait}

In FIG. \ref{fig1} the phase portraits of the dynamical system \eqref{ode-1} for toy model 1 (top panels) and of \eqref{ode-mod2} for toy model 2 (bottom panels), are shown. The critical points $P_\text{bb}$ (exists only for vanishing coupling $\lambda_k=0$) and $P_\text{mat}$ appear enclosed by the small circles, while the de Sitter attractor manifold ${\cal M}_\text{dS}$ is represented by the dash-dot line joining the points $(0,1)$ and $(1,1)$. This manifold coincides with the upper boundary of the phase square. The inflationary quantum manifold ${\cal Q}_\text{infl}$ is represented by the dash-dot line with coordinate $y=0$, starting at the origin and ending up at the solid circle with coordinate $\left\langle x\right\rangle$ (it coincides with the corresponding segment of the lower boundary). As a matter of fact $\left\langle x\right\rangle$ is an averaged value due to quantum fluctuations of the boundary of the inflationary manifold. The value $\left\langle x\right\rangle$ defines the condition $H\approx L^{-1}_\text{Pl}$, i. e., it represents the fuzzy boundary of the quantum domain inside which the equations of the present formalism are not valid anymore. The straight vertical dot line represents the boundary where $H=L^{-1}$. To the left of this boundary the energies are higher than the new scale $L^{-1}$.

For completeness, the relevant curves $\bar y_j=\bar y_j(x)$ ($j=1$ for model 1 while $j=2$ for model 2) in \eqref{sep-y} and \eqref{sep-mod2}, respectively, i .e., the separatrices represented by the solid dark curves in the figure, and $\hat y_j=\hat y_j(x)$ in \eqref{q0} and \eqref{q0-mod2}, respectively (dash dark curves), have been included as well in the phase portraits. The region below the separatrix -- gray shading -- corresponding to anti-de Sitter background spaces, is not of interest for the present investigation. The region with $\bar y_j\leq y\leq\hat y_j$ (magenta shading in the figure) is where the expansion happens at an accelerated pace.

%-------------------------------------------------------------------------------

\subsubsection{$\Lambda$CDM model}\label{subsect-lcdm}

The existence of the energy scale $L^{-1}\lesssim L^{-1}_\text{Pl}$ appreciably modifies the global dynamics of the $\Lambda$CDM model emerging from the geometric inflation formalism when compared with the known GR-based result. Actually, the $\Lambda$CDM model retrieved from the present set up in the limit when the coupling $\lambda_k\rightarrow 0$, does not exactly coincide with the one obtained within the framework of general relativity. In particular, the quantum manifold ${\cal Q}_\text{infl}$ is replaced by the global past attractor $P_\text{bb}:(0,1/2)$. This latter equilibrium state has no analogue in the GR-based $\Lambda$CDM model where only two critical points can be found: i) the matter-dominated past attractor and ii) the de Sitter future attractor. This discrepancy is due to the fact that the point $P_\text{bb}$ is deep inside the high energy domain $H\gg L^{-1}$ (vertical dot line in FIG. \ref{fig1}) while the GR-based $\Lambda$CDM model is for energies $H\ll L^{-1}$, where the effects of the new scale have no impact in the cosmological dynamics. As a matter of fact, the limit $\lambda_k\rightarrow 0$ is equivalent to the ``classical limit'' $L_\text{Pl}\rightarrow 0$. Hence, this limit may be thought of as the classical limit of the geometric inflation formalism, where the quantum effects are neglected. This is why the bigbang cosmological singularity arises in this limit.

In the $\lambda_k\rightarrow 0$ limit of the present formalism the matter-dominated solution $P_\text{mat}$ is a saddle point. It attracts orbits along the $x$ direction that are sourced at $P_\text{bb}$ and then repels them along the $y$ direction towards a de Sitter point in ${\cal M}_\text{dS}$. This is why one of the eigenvalues $\lambda_1=-3(\omega_m+1)/2$ (the one related with the eigenvector along the $x$ direction) is negative, while the other one $\lambda_2=3(\omega_m+1)/2$, which is related with the eigenvector along the $y$ direction, is positive. In the low curvature limit $H^2\ll L^{-2}$ ($x\rightarrow 1$ $\Rightarrow LH\rightarrow 0$), the $x$ dimension is shrank to a point and we are left with a phase line $\{(x,y):x=1,1/2\leq y\leq 1\}$. In this limit only the positive eigenvalue remains so that the point $P_\text{mat}$ transforms into the global past attractor. This result shows that the correct GR-limit of the geometric inflation set up is $LH\rightarrow 0$, instead of $\lambda_k\rightarrow 0$. We shall discuss more on this in section \ref{sect-discuss}.

It is worth noticing that the $\Lambda$CDM limit of the geometric inflation formalism is a classic limit in the sense that the coupling $\lambda_k$ of the higher order curvature contributions to gravity is vanishing. This means that any geometrical manifestation of the quantum effects is eliminated. This is why in the left-hand panels of FIG. \ref{fig1} for the $\Lambda$CDM limit of the geometric inflation model, the quantum domain (dark-gray vertical strip) is not visible. As a matter of fact, in the $\lambda_k\rightarrow 0$ limit of the geometric inflation formalism, the quantum domain is shrank to the neighborhood of the point $x=0$, i. e., in this case $L^{-1}\lesssim L^{-1}_\text{Pl}\rightarrow\infty$. However, as commented above, even in this classic limit the existence of the scale $L^{-1}$ makes a difference as compared with general relativity.

%-------------------------------------------------------------------

\subsubsection{Quantum inflationary manifold ${\cal Q}_\text{infl}$}

As said this is a fuzzy manifold since it falls within the quantum domain $H\geq L_\text{Pl}^{-1}$. Along the manifold, from $x=0$ to $x=\left\langle x\right\rangle$, the Hubble parameter changes from very high curvature regime $H\gg L^{-1}$ at $x=0$, to 

\bea H({\left\langle x\right\rangle})=\sqrt\frac{1-\left\langle x\right\rangle}{\left\langle x\right\rangle}\,L^{-1},\label{hxav}\eea at $x=\left\langle x\right\rangle$. Although an analytical expression for $\left\langle x\right\rangle$ can not be found, it can be evaluated numerically. In FIG. \ref{fig1} $\left\langle x\right\rangle$ is represented by the solid circle. It is determined by the chosen criterion according to which quantities with values $\leq 10^{-5}\lambda_k$ may be assumed vanishing with an appropriate level of accuracy. Hence, for the value $\lambda_k=10^{-5}$, where $k=4$ for model 1 while $k=3$ for model 2 (panels in the left-hand column in FIG. \ref{fig0}) we obtain that $\left\langle x\right\rangle=0.17$ for model 1 while $\left\langle x\right\rangle=0.05$ for model 2. For $\lambda_k=10^{-2}$ (panels in the middle column of FIG. \ref{fig0}) we get that $\left\langle x\right\rangle=0.22$ for model 1 and $\left\langle x\right\rangle=0.1$ for model 2. For panels in the right-hand column of FIG. \ref{fig0}, $\lambda_k=1$ so that $\left\langle x\right\rangle=0.3$ for model 1 while $\left\langle x\right\rangle=0.2$ for model 2. 

In this paper we identify the averaged value $\left\langle x\right\rangle$ with the position of the quantum boundary in the phase portrait, i. e., with the point where $H\approx L^{-1}_\text{Pl}$. Since, according to \eqref{hxav}, in this case: $$H=L^{-1}\sqrt\frac{1-\left\langle x\right\rangle}{\left\langle x\right\rangle}\approx L^{-1}_\text{Pl},$$ then

\bea L^{-1}\approx\sqrt\frac{\left\langle x\right\rangle}{1-\left\langle x\right\rangle}L^{-1}_\text{Pl}.\label{rel}\eea Hence, for the choice $\lambda_4=10^{-5}$ in model 1, since $\left\langle x\right\rangle=0.17$, we get $L^{-1}\approx 0.45 L^{-1}_\text{Pl}$. If, on the contrary, one establishes a priori a relationship between the new and the Planck scales, for instance: $L^{-1}\approx 10^{-1} L^{-1}_\text{Pl}$, this leads to $\left\langle x\right\rangle\approx 10^{-2}$. At this point $dx/dv\approx 10^{-4269}=10^{-4264}\times 10^{-5}\lambda_4$, which is much much below the established criterion based on the quantity $10^{-5}\lambda_4$. This means that the relationship between the new and the Planck scales is what determines the accuracy in the identification of the inflationary quantum manifold.

It is interesting to note that orbits that originate deep inside the quantum region are vertical lines joining a point in ${\cal Q}_\text{infl}$ with a point with same $x$-coordinate in the attractor de Sitter manifold ${\cal M}_\text{dS}$. This means that initial conditions deep inside the quantum domain lead to de Sitter ever expanding sterile universes (cosmic structure is not formed at any stage) with constant $$H=H_0=\sqrt\frac{1-x_0}{x_0}\,L^{-1},$$ where $x=x_0$ is the initial condition. We have to recall, however, that inside the quantum domain we may not trust the results obtained on the basis of the present classical theory.

In order for a given orbit to lead to sensible cosmic dynamics it should leave the quantum domain and after a primordial inflationary period to meet the region of the phase space where the expansion occurs at a decelerated peace, i. e., the region where $\bar y_j(x)\leq y<\hat y_j(x)$, with $\bar y_j(x)$ and $\hat y_j(x)$ ($j=1,2$) given by \eqref{sep-y}, \eqref{sep-mod2} and \eqref{q0}, \eqref{q0-mod2}, respectively. In FIG. \ref{fig1} the orbits that lead to well-behaved cosmic dynamics are same color than the magenta-shading region where the deceleration parameter $q>0$. The decelerated expansion stage is mandatory for the required amount of cosmic structure to form. For those orbits that evolve always in the region where the expansion is accelerated (black thin solid curves) no cosmic structure forms at all. In consequence we call this region as sterile inflation region.

%--------------------------------------------------------------------

\subsubsection{de Sitter attractor manifold ${\cal M}_\text{dS}$}

The de Sitter attractor manifold owes its existence to the non-vanishing cosmological constant $\Lambda$ that is included in the model from the start \cite{arciniega-2}. This warrants that the decelerated expansion, whenever it takes place, can be only a transient stage of the cosmic evolution. From FIG. \ref{fig1} it is seen that those orbits that represent sensible cosmic dynamics start in a stage of primordial curvature inflation, then go into a stage of decelerated expansion where the appropriate amount of cosmic structure forms (magenta-shading region in FIG. \ref{fig1}), to finally enter another inflationary and, eventually, end up at a de Sitter regime.

%%%%%%%%%%%%%%%%%%%%%%%%%%%%%%%%%%%%%%%%%

\section{Discussion}\label{sect-discuss}

One of the most important features of the geometric inflation formalism \cite{arciniega-2} is the existence of the energy scale $L^{-1}\lesssim L^{-1}_\text{Pl}$. This property is independent of the specific model considered to sum the infinite series in \eqref{action}.

The importance of the new energy scale $L^{-1}$ is clearly illustrated in the limit $\lambda_k\rightarrow 0$ of the models, when the $\Lambda$CDM paradigm is retrieved from the present set up. As stated in subsection \ref{subsect-lcdm}, in this limit the global dynamics differs from the one obtained in the GR-based model. For the GR-based $\Lambda$CDM model the motion equations read:

\bea &&\Omega_m+\Omega_\Lambda=1,\nonumber\\
&&2\frac{\dot H}{H^2}=-3(w_m+1)\Omega_m,\nonumber\\
&&\rho'_m=-3(w_m+1)\rho_m,\label{lcdm-moteq}\eea where the prime denotes derivative in respect to the number of e-foldings $\tau=\ln a$. Given the Friedmann constraint (first equation above), in this case only one of the dimensionless energy density parameters can be an independent variable of some phase line. Take, for instance, $0\leq\Omega_m\leq 1$ to be the independent variable. We can trade the cosmological equations \eqref{lcdm-moteq} by the following autonomous ordinary differential equation:

\bea \Omega'_m=-3(w_m+1)\Omega_m\left(1-\Omega_m\right).\label{ode-lcdm}\eea The critical points of this one-dimensional dynamical system are: i) $\Omega_m=1$, the matter-dominated source point and ii) $\Omega_m=0$, the de Sitter attractor point. This is to be contrasted with the $\Lambda$CDM limit, $\lambda_k\rightarrow 0$, of the present formalism (see the discussion in subsection \ref{subsect-lcdm} and the left-hand panels of FIG. \ref{fig1}). In this limit there are two critical points and a critical manifold. The past global attractor is the matter-dominated bigbang $P_\text{bb}:(0,1/2)$, where $\Omega_m=1$ and, since $x\rightarrow 0$ then $H\gg L^{-1}$. This solution has no analogue in the GR-based $\Lambda$CDM model. In addition to the former equilibrium point there is the saddle standard or GR matter-dominated solution $P_\text{mat}:(1,1/2)$ where $3H^2=\rho_m$. In this case since $x\rightarrow 1$, then $H\ll L^{-1}$. This is to be contrasted with the similar solution in the GR-based $\Lambda$CDM model which is the past attractor instead of a saddle point in the phase plane. Finally, in the present set up there is the de Sitter manifold ${\cal M}_\text{dS}$ \eqref{mds} constituted by attractor de Sitter points characterized by $$H=H_0=L^{-1}\left(\frac{1-x_0}{x_0}\right),$$ with $x_0$ being a fixed value in the interval $0\leq x_0\leq 1$. In the left-hand end of the interval, i. e., in the $x_0\rightarrow 0$ limit, we have de Sitter expansion with very large $H_0\gg L^{-1}$ while, in the right-hand end where $x_0\rightarrow 1$, we get GR de Sitter expansion with $H_0\ll L^{-1}$. Critical points in this limit ($x\rightarrow 1$) are the ones analogous to the standard de Sitter attractor in the GR-based $\Lambda$CDM model.

As seen from the above discussion, in order to get the standard GR-based $\Lambda$CDM model as a particular case of the geometric inflation set up, it is not necessary (not enough) to take the limit $\lambda_k\rightarrow 0$ but, instead, it is required to go to the low-curvature limit $LH\rightarrow 0$ ($H\ll L^{-1}$). In this limit the critical point $P_\text{bb}:(0,1/2)$ does not exists and one of the eigenvalues $\lambda_\pm=\pm 3(w_m+1)/2$ of the linarization matrix for the matter-dominated point $P_\text{mat}:(1,1/2)$, vanishes since the phase space decreases dimension from 2 to 1 in this limit: $\Psi_\text{phys}$ shrinks to the vertical line at $x=1$. Only the positive eigenvalue $\lambda_+$ survives rendering $P_\text{mat}$ the source point as it is for the GR-based situation.

%=========================================================================

\subsection{On the inflationary stages in the present set up}

In what regards to the inflationary stages in the geometric inflation model, there are two of them. One primordial inflationary period and a second inflationary stage at late times. For some orbits -- those leading to sterile evolution -- these stages are continuously joined while for others -- the ones that allow for the correct amount of cosmic structure to form -- an intermediate epoch of decelerated expansion joins the two inflationary stages (see FIG. \ref{fig1}). Hence, the arising of a sensible cosmological dynamics in the geometric inflation model depends on the initial conditions.

From the motion equations \eqref{moteq-m1} and \eqref{moteq-m2} it follows that

\bea -2\dot H=\frac{(w_m+1)\rho_m}{1+2\lambda_4(2+L^4H^4)L^6H^6\,e^{L^4H^4}},\label{doth-m1}\eea for the model 1, while

\bea -2\dot H=\frac{(w_m+1)\rho_m}{1-3\lambda_3L^4H^4+\lambda_3(4+L^2H^2)L^6H^6\,e^{L^2H^2}},\label{doth-m2}\eea for model 2, where $\rho_m\propto a^{-3(w_m+1)}$. In the very high curvature limit $H\gg L^{-1}$, the de Sitter expansion is approached as long as $$L^{10}H^{10}e^{L^4H^4}\gg a^{-3(w_m+1)},$$ for the model 1 or $$L^8H^8e^{L^2H^2}\gg a^{-3(w_m+1)},$$ for model 2, so that $\dot H\rightarrow 0$. But since $L^{-1}\lesssim L^{-1}_\text{Pl}$, this means that the very high curvature regime $H\gg L^{-1}$ falls in the quantum domain where the motion equations of the geometric inflation model stop being valid. Hence, while at late times the expansion is de Sitter $a(t)\propto\exp(\sqrt{\Lambda/3}t)$, at early times the inflationary stage is not always de Sitter. As we have just shown, the primordial de Sitter inflation takes place in the quantum domain (vertical orbits in the dark gray-shading strip in FIG. \ref{fig1}) so that, no sensible cosmic dynamics can be linked with it. Besides, within the quantum domain the motion equations of the present formalism are not valid anymore, so that it is not clear whether the primordial de Sitter expansion actually takes place in this set up. 

The sensible cosmological scenario in the present formalism takes place if for large (but not too much) curvature $L^{-2}\lesssim H^2\lesssim L^{-2}_\text{Pl}$, i. e., for cosmic evolution in the part of the phase square in FIG. \ref{fig1} to the left of the vertical dot line and without the dark-gray strip representing the quantum domain, the crossing of the condition $q=0$ in the direction from $q<0$ to $q>0$, is possible. For the choice $w_m=1/3$ (background radiation), at early times where $\rho_m\gg\Lambda$ (i. e., one may safely set $\Lambda=0$), the deceleration parameter $q=-1-\dot H/H^2$ reads

\bea q=\frac{1-2\lambda_4(1+L^4H^4)L^6H^6e^{L^4H^4}}{1+2\lambda_4(2+L^4H^4)L^6H^6e^{L^4H^4}},\label{q-mod-1}\eea for model 1, while for model 2 we have:

\bea q=\frac{1+\lambda_3L^4H^4-\lambda_3(2+L^2H^2)L^6H^6e^{L^2H^2}}{1-3\lambda_3L^4H^4+\lambda_3(4+L^2H^2)L^6H^6e^{L^2H^2}}.\label{q-mod-2}\eea Hence, the condition $q=0$ amounts to:

\bea (1+L^4H^4)L^6H^6e^{L^4H^4}=\frac{1}{2\lambda_4},\label{q1-mod-1}\eea for model 1 and 

\bea (2+L^2H^2)L^6H^6e^{L^2H^2}=\frac{1}{\lambda_3}+L^4H^4,\label{q1-mod-2}\eea for model 2. Whenever the left-hand side (LHS) in the above equations is greater than the right-hand side (RHS) -- recalling that $L^{-1}\lesssim H\lesssim L^{-1}_\text{Pl}$ --, primordial non-de Sitter inflation takes place. Then, as long as $H$ further decreases so that the LHS of equations \eqref{q1-mod-1} and \eqref{q1-mod-2} becomes smaller than the RHS, the crossing of the condition $q=0$ takes place and the corresponding phase space orbits enter a decelerated expansion region (magenta shading in FIG. \ref{fig1}), with the consequent formation of cosmic structure. Notice from equations \eqref{q-mod-1} and \eqref{q-mod-2} that, if consider the very high curvature regime $H^2\gg L^{-2}$ $\Rightarrow H^2>L^{-2}_\text{Pl}$, it follows that $q\rightarrow-1$, i. e., this is a de Sitter expansion regime taking place in the quantum domain $H^{-1}<L_\text{Pl}$, as discussed above.

%%%%%%%%%%%%%%%%%%%%%%%%%%%%%%%%%%%%%%%%%%%%

\section{Conclusion}\label{sect-conclu}

In this paper, based on the dynamical systems analysis, we have discussed on the geometric origin of primordial inflation in the geometric inflation formalism \cite{arciniega-2}. We have investigated two toy models proposed in the mentioned reference, where the sum over the infinite tower of higher-order curvature invariants is performed, yielding to compact expressions in the equations of motion. These are very encouraging possibilities if regard gravity as a quantum effective theory \cite{donoghue}, since higher powers of $R$, $R_{\mu\nu}$ and $R_{\mu\nu\sigma\lambda}$, would be involved at higher loops. Hence, one would naively expect that consideration of the whole infinite tower of curvature invariants, would amount to consideration of all of the higher order loops, so that quantum effects would be manifest.

Perhaps the more interesting result of the present research has been to show, precisely, the quantum origin of the primordial inflation in the geometric inflation theory. This is a consequence of the new length scale $L$ which is assumed above (but not too much) of the Planck length $L_\text{Pl}$: $L\gtrsim L_\text{Pl}$. As seen from FIG. \ref{fig1}, for equilibrium points in the inflationary past attractor we have that $H^{-1}<L_\text{Pl}$, so that any orbit that leads to either unphysical or sensible cosmic dynamics, starts in the past attractor (or in its boundary) within the quantum domain.

In order to allow for under-Planckian initial energy densities, recently an hybrid geometric inflation model was proposed \cite{edelstein}, where the role of a scalar field in the geometric inflation formalism is investigated. Although this model misses one of the most attractive features of the geometric inflation formalism: the pure geometrical origin of primordial inflation, it would be very interesting to explore the asymptotic dynamics of such an hybrid scenario. This will be the subject of forthcoming work.

%%%%%%%%%%%%%%%%%%%%%%%%%%%%%%%%%%%%%%%%%%%%%%

\section{Acknowledgments}

The authors are grateful to SNI-CONACyT for continuous support of their research activity. The work of RGS was partially supported by SIP20200666, COFAA-IPN, and EDI-IPN grants. FXLC also acknowledges the Programa para el Desarrollo Profesional Docente (PRODEP) for financial support. UN thanks PRODEP-SEP and CIC-UMSNH for financial support of his contribution to the present research. 

%%%%%%%%%%%%%%%%%%%%%%%%%%%%%%%%%%%%%%%%%%%%%%%%%%%%%%%

%%%%%%%%%%%%%%%%%%%%%%%%%%%%

\end{document}